\documentclass[preprint,journal]{vgtc}       

\ifpdf
  \pdfoutput=1\relax                   
  \usepackage{graphicx}                
  \DeclareGraphicsExtensions{.pdf,.png,.jpg,.jpeg} 
\else
  \usepackage{graphicx}                
  \DeclareGraphicsExtensions{.eps}     
\fi%

\graphicspath{{figures/}{pictures/}{images/}{./}} 

\usepackage{microtype}                 
\PassOptionsToPackage{warn}{textcomp}  
\usepackage{textcomp}                  
\usepackage{mathptmx}                  
\usepackage{times}                     
\usepackage{cite}                      
\usepackage{tabu}                      
\usepackage{booktabs}                  

\usepackage{multirow} 

\usepackage{caption}
\usepackage{subcaption}

\usepackage{amsmath}
\DeclareMathOperator*{\argmin}{argmin}
\usepackage{amsthm}
\usepackage{commath}                  
\usepackage{bm} 

\theoremstyle{remark}

\usepackage{xcolor} 
\usepackage[normalem]{ulem}
\newcommand{\revision}[1]{\textcolor{blue}{#1}}
\newcommand{\removed}[1]{\textcolor{red}{\sout{#1}}}

\def \cleanversion{} 
\ifx\cleanversion\undefined
\else
  \renewcommand{\revision}{}
  \renewcommand{\removed}[1]{\iffalse#1\fi}
\fi

\onlineid{1025}

\vgtccategory{Representations \& Interaction}
\vgtcpapertype{algorithm/technique}

\title{VMap: An Interactive Rectangular Space-filling Visualization for Map-like Vertex-centric Graph Exploration}

\author{Jiayi~Xu and Han-Wei~Shen}
\authorfooter{
\item
 Jiayi Xu is with the Department of Computer Science and Engineering, The Ohio State University. E-mail: xu.2205@osu.edu.
\item
 Han-Wei~Shen is with the Department of Computer Science and Engineering, The Ohio State University. E-mail: shen.94@osu.edu.
}

\shortauthortitle{Xu \MakeLowercase{\textit{et al.}}: VMap: An Interactive Rectangular Space-filling Visualization for Map-like Vertex-centric Graph Exploration}

\abstract{
We present VMap, a map-like rectangular space-filling visualization, to perform vertex-centric graph exploration. Existing visualizations have limited support for quality optimization among rectangular aspect ratios, vertex-edge intersection, and data encoding accuracy. 
To tackle this problem, VMap integrates three novel components: (1) a desired-aspect-ratio (DAR) rectangular partitioning algorithm, (2) a two-stage rectangle adjustment algorithm, and (3) a simulated annealing based heuristic optimizer. 
First, to generate a rectangular space-filling layout of an input graph, we subdivide the 2D embedding of the graph into rectangles with optimization of rectangles' aspect ratios toward a desired aspect ratio. 
Second, to route graph edges between rectangles without vertex-edge occlusion, we devise a two-stage algorithm to adjust a rectangular layout to insert border space between rectangles. 
Third, to produce and arrange rectangles by considering multiple visual criteria, we design a simulated annealing based heuristic optimization to adjust vertices' 2D embedding to support trade-offs among aspect ratio quality and the encoding accuracy of vertices' weights and adjacency. 
We evaluated the effectiveness of VMap on both synthetic and application datasets. The resulting rectangular layout has better aspect ratio quality on synthetic data compared with the existing method for the rectangular partitioning of 2D points. On three real-world datasets, VMap achieved better encoding accuracy and attained faster generation speed compared with existing methods on graphs' rectangular layout generation. We further illustrate the usefulness of VMap for vertex-centric graph exploration through three case studies on visualizing social networks, representing academic communities, and displaying geographic information. 
} 


\CCScatlist{ 
 \CCScat{K.6.1}{Management of Computing and Information Systems}%
{Project and People Management}{Life Cycle};
 \CCScat{K.7.m}{The Computing Profession}{Miscellaneous}{Ethics}
}

\vgtcinsertpkg

\begin{document}


\firstsection{Introduction}

\maketitle

\revision{\textit{Vertex-centric} (or, \textit{node-centric}) graph exploration refers to first discovering vertices of interest and then navigating over graphs interactively. 
For example, the social network analysis~\cite{zhao2016egocentric, chen2016d} can start with the identification of influential persons and continue with the exploration of networks centered on these opinion leaders. 
The transport network analysis~\cite{zeng2014visualizing} can start with the identification of locations of interest and continue with the exploration of possible origin-destination routes between these locations. Other vertex-centric graph exploration applications include research collaboration networks~\cite{gansner2010gmap, wu2015egoslider, zhao2016egocentric}, resource description networks~\cite{sayers2004node, schlegel2014balloon}, co-purchasing networks in Amazon~\cite{gansner2010gmap}, and topic networks in Wikipedia~\cite{biuk2014visualizing}, just to name a few. }



\revision{
Map-like visualization for vertex-centric graphs separates the entire display space into separate regions that are then allocated to represent different vertices, utilizing the cognitive benefits of cartographic maps as explained by a survey~\cite{hografer2020state}. 
Studies on map-like vertex-centric graph exploration
mainly focus on three graph entities: (1) clusters of vertices, (2) specific vertices of interest, and (3) connections centered on vertices of interest. 
Correspondingly, we identify three levels of vertex-centric graph exploration. 
First, in cluster-level exploration, vertices with high proximity are grouped together, enabling analysis of group functions and efficient hierarchical exploration. 
Second, in vertex-level exploration, information attached to vertices of interest is displayed, enabling analysis of individual attributes and vertices' diversity. 
Third, in connection-level exploration, ego-centric connections centered on vertices of interest are revealed and investigated, enabling in-depth analysis of individual behaviors and relationships. }

%

%

\revision{Challenges exist to design a map-like visualization to perform the three levels of vertex-centric graph exploration. 
First, at the cluster level, it is important to arrange vertices on the same clusters together to minimize confusion; however, this is not guaranteed by certain existing visualizations, such as~\cite{gansner2010gmap, chen2016d}. 
Second, at the vertex level, it is essential to faithfully encode various information attached to vertices, including numerical attributes, derived metrics (e.g., node-centrality metrics), and text labels. However, existing visualizations either cannot guarantee the faithful encoding of numeric values or have limited power to make trade-offs between different encodings. 
Third, at the connection level, it is crucial to support ego-centric exploration and arrange edges with minimal visual clutter. 
To the best of our knowledge, it is still an open problem for visualization researchers to solve the three challenges. }

\revision{
We design an interactive VMap (short for Vertex Map) to tackle the aforementioned three challenges of vertex-centric graph visualization. 
On a VMap, similar to a rectangular treemap, each vertex is represented by a rectangle, vertices' numerical information is encoded using area, vertices' labels can be embedded within rectangles, and the nesting structure of rectangles encodes the grouping of vertices. 
Compared with the treemap-based graph studies~\cite{fekete2003interactive, holten2006hierarchical}, VMap has two main distinctions: (1) vertices' layout optimized for vertices' connections and (2) edges' placement with the use of border space of rectangles to avoid visual clutter. 
To construct a VMap, we design a visualization framework with three components (1) a Desired-Aspect-Ratio (DAR) technique for the rectangular layout of vertices and (2) a two-stage rectangle adjustment algorithm for placement of edges, and (3) a simulated annealing optimization to improve final results, and the three components work hand-in-hand to produce satisfied visualizations. 
First, the DAR approach guarantees vertices within the same cluster are contained by the same high-level rectangle, and the rectangles' areas faithfully encode vertices' numerical attributes. DAR enhances the binary space partitioning method by a DAR heuristic to arrange rectangles with satisfied aspect ratios. 
Second, the two-stage adjustment algorithm is proposed to excavate the border space between rectangles for edge placement, so that users can explore ego-centric networks interactively with minimized visual clutter. The adjustment algorithm post-processes resulting rectangular layouts of DAR, in the meantime, protect small rectangles from being eliminated and preserve areal proportions of rectangles. 
Third, the simulated annealing based optimization is designed to improve the rectangular layout by considering three aspects: vertices' connections, encoding faithfulness, and rectangles' aspect ratio, and also supports generating layouts with trade-offs among the three aspects. 
Hence, our contributions are threefold: }
\revision{\begin{enumerate} 
\item We propose a DAR binary partitioning algorithm to produce vertices' rectangles with low aspect ratio loss. 
\item To excavate space for edges, we offer a two-stage rectangle adjustment technique that guarantees areal proportion preservation to insert borders with a fixed width. 
\item We propose simulated annealing based heuristic optimization to optimize the final rectangular layout and make trade-offs among different areal error, topological error, and aspect ratio loss.  
\end{enumerate}}

\section{Related Works}





We discuss related techniques regarding vertex-centric graph visualizations with map metaphor in the following. 
\subsection{Vertex-centric Graph Visualization with Map Metaphor}
We classify existing map-like vertex-centric graph visualizations into three categories: (1) Voronoi-diagram-based~\cite{gansner2010gmap}, (2) force-directed-adjusting-based~\cite{biuk2014visualizing, efrat2014mapsets, chen2016d}, (3) hierarchy-based~\cite{fekete2003interactive, holten2006hierarchical, gronemann2012drawing, auber2013gospermap}, and (4) rectangular-dual-based~\cite{he1993finding, van2007rectangular, speckmann2006linear, buchin2010optimizing, buchin2012evolution, buchin2011adjacency, carrizosa2018mathematical} visual representations according to the method used for space partitioning, where our DAR method is binary-space-partitioning (BSP) based and is different from all the existing methods for graphs. We also demonstrate the limitations of existing methods on vertex encoding and placement in the following. 


\textbf{Voronoi-based visualization:} 
After obtaining a 2D embedding of a given graph, Gansner et al.~\cite{gansner2010gmap} constructed GMap by using the Voronoi diagram to partition the 2D embedding of vertices and assign each vertex a Voronoi cell, where distinct colors encode clusters of vertices. 
GMap has limitations, including that (1) vertices' Voronoi cells in the same cluster may be noncontiguous, and (2) GMap does not directly support encoding vertices' numerical attributes. 

\textbf{Force-directed-adjusting-based visualization:} 
The visualizations are generated by representing each vertex using a unit convex shape and adjusting vertices' positions using spring-like forces~\cite{eades1984heuristic, kamada1989algorithm, fruchterman1991graph} to satisfy certain criteria. 
D-Map proposed by Chen et al.~\cite{chen2016d} merges vertices' unit hexagons to make a compact visual representation by adding forces to push vertices in the same cluster closer and toward the central vertex. Similar to GMap, clusters of vertices in D-Map are visualized by distinct colors. 
Limitations of D-Map include that (1) vertices' hexagons in the same cluster may be isolated, and (2) D-Map is not designed to support encoding vertices' numerical attributes. 
To ensure geometries of vertices in the same cluster are contiguous, Biuk-Aghai et al.~\cite{biuk2014visualizing} used a force transfer technique~\cite{huang2003force} and Efrat et al.~\cite{efrat2014mapsets} designed MapSets using a force-directed adjustment. These two methods~\cite{biuk2014visualizing, efrat2014mapsets} rely on extracting trees from input graphs to represent clusters and preserve topology, and hence have a limitation that most edges are not paid attention for vertex placement, and connected vertices may not be near enough. 
A limitation of force-directed-adjusting based techniques~\cite{biuk2014visualizing, efrat2014mapsets, chen2016d} in general is that these methods may leave blank in display space and may not generate space-filling visualizations, causing space usage not efficient enough.

\textbf{Hierarchy-based visualization:} 
Treemaps are originally designed to represent hierarchical data using space-filling rectangles~\cite{Shneiderman1992, johnson1991tree}, and used to represent hierarchical trees extracted from input graphs that are hierarchical and vertex-weighted. 
Fekete et al.~\cite{fekete2003interactive} visualized hierarchical trees extracted from hierarchical file systems and websites' XML documents. Holten~\cite{holten2006hierarchical} visualized hierarchical source code trees extracted from call relationships between source codes. 
Afterwards, rather than just rectangles, treemaps are extended to represent hierarchy information by convex polygons~\cite{de2010fat, de2014treemaps}, which are then also utilized to visualize hierarchical weighted binary trees extracted from graphs by~\cite{gronemann2012drawing}. 
GosperMap~\cite{auber2013gospermap} also supports visualizing hierarchy information extracted from graphs. 
Following this idea, other map-like visualizations for hierarchy data, such as~\cite{skupin2004world, biuk2013novel}, can also have the potential for vertex-centric graph visualizations in the future, although those have not been studied by researchers yet. 
The limitations of hierarchy-based graph visualizations include that (1) input graphs are restricted to contain meaningful hierarchy information, and (2) vertices' placement uses limited connectivity information leading to strongly-connected vertices may be placed far away.

\textbf{Rectangular-dual based visualization:} 
Given a planar graph with certain topology constraints, rectangular dual based methods~\cite{he1993finding} generate a rectangular subdivision, where every vertex is represented by a rectangle and edges are expressed by contacts of related rectangles; note that the vertex and rectangle are bijective correspondence as well as the edge and contact of rectangles. The traditional rectangular dual methods are not designed to encode vertices' numerical attributes and have limitations to represent regions' statistical quantities. To improve that, rectangular-dual-based cartograms are proposed. 

The concept of the rectangular dual is extended to draw rectangular cartograms. The adjacency of geographic regions can be transformed into a planar graph, on which a vertex represents a region and an edge represents whether two regions are contiguous. 
To adjust rectangles to represent vertices' statistical quantities by areas of rectangles approximately, Van Kreveld et al.~\cite{van2007rectangular, speckmann2006linear} used linear programming and Buchin et al. proposed simulated annealing based~\cite{buchin2010optimizing} and evolution based~\cite{buchin2012evolution} approaches. 
Moreover, to extend the previous works for spatial data with two hierarchy levels of regions, Buchin et al.~\cite{buchin2011adjacency} proposed a post-processing method called adjacency-preserving spatial treemaps. The rectangular-dual-based cartogram techniques require that input graphs have planar 2D embeddings and every vertex has a geographic location, and hence have limitations that cannot directly take general graphs as input.

\subsection{Edge Encoding and Placement}\label{sect:related_works_topology}

Rather than without edge encodings in~\cite{auber2013gospermap, biuk2014visualizing, efrat2014mapsets}, most existing studies visualize edges by (1) overlaying edges on map-like visualizations and (2) using shape contact encodings. Note that no existing studies use border space to route edges similar to our VMap. 

\textbf{Overlay:} 
Edges can be placed upon map-like visualizations, such as in~\cite{fekete2003interactive, holten2006hierarchical, gansner2010gmap, gronemann2012drawing, chen2016d}. 
To reduce occlusion and visual clutter caused by edge encodings, researchers usually choose to use edge bundling techniques~\cite{holten2006hierarchical, gronemann2012drawing} or display a subset of edges in response to users' interactions~\cite{chen2016d}. 

\textbf{Shape contact:} 
Edges are encoded using rectangle contacts in rectangular-dual-based graph visualizations. The traditional rectangular-dual methods~\cite{he1993finding} are limited for certain types of planar graphs. 
To satisfy the topology restrictions of rectangular-dual methods, Van Kreveld et al.~\cite{van2007rectangular} and Speckmann et al.~\cite{speckmann2006linear} introduced ``sea'' rectangles to help layout arrangements, which also ensures the union of vertices' rectangles and sea rectangles is still a rectangle. 
The ``sea'' in~\cite{van2007rectangular, speckmann2006linear, buchin2010optimizing, buchin2012evolution} is a meaningful metaphor for cartograms, however, it becomes blank space when visualizing general graphs, reducing overall space usage. 
To optimize the rectangular layout for a general graph without introducing ``sea'', Carrizosa et al.~\cite{carrizosa2018mathematical} allowed certain topological errors and relaxed the constraint of the bijection between edges and rectangle contacts. 
The method of Carrizosa et al.~\cite{carrizosa2018mathematical} has three steps. First, input graphs are embedded into a two-dimensional space using the multi-dimensional scaling (MDS) method. Second, the embedding is optimized by an objective function with twenty-six optimization constraints. Third, Carrizosa et al. introduced a cell perturbing algorithm to find grid points of vertices and rectangular regions for vertices on a coarse grid ($10 \times 10$ by default). An embedded cell perturbing algorithm was proposed to extend the solution on a coarse grid to a tighter grid ($20 \times 20$ by default). The limitations of~\cite{carrizosa2018mathematical} include that (1) twenty-six complex optimization constraints make optimization time long and (2) optimization of aspect ratios of rectangles is not supported, while non-extreme aspect ratios help users perceive the weights of vertices well \cite{kong2010perceptual} and offer appropriate space to display labels of vertices clearly.

\section{VMap Overview}
We provide the framework overview of VMap. 
\revision{The input of VMap is a \textit{vertex-weighted graph} $G = (V, E, \alpha)$. $V$ is the set of vertices, and $E$ is the set of edges. For a vertex $v_i \in V$, we define its weight proportion to be $\alpha_i$. }
\revision{
The output is a rectangular space-filling visualization $G^P = (V^P, E^P, \alpha^P, w, h)$. Rectangles in $V^P$ and vertices in $V$ have a bijective correspondence. $E^P$ is the set of rectangle contacts, where two rectangles have a contact if and only if their intersection is not empty, namely, which is a point or a segment. For a rectangle $v^P_i \in V^P$, its \textit{area proportion} is $\alpha^P_i$, its \textit{width} is $w_i$, and its \textit{height} is $h_i$. }

\textbf{Method pipeline:} 
We initialize the 2D embedding of vertices randomly and later optimize the embedding based on a simulated annealing method. Given a vertices' 2D embedding, we generate space-filling rectangles using the DAR rectangular partitioning. We arrange rectangles for clusters first, and then subdivide clusters' rectangles to enclose rectangles of vertices within clusters. 
When DAR partitioning is used, the aspect ratio of each rectangle is targeted at a desired number ($1.5$ by default according to previous perception studies~\cite{kong2010perceptual, heer2010crowdsourcing}), and zero areal error is guaranteed. 
By using a two-stage rectangle adjustment, we open a border space between rectangles to route edges amongst vertices, and the visual elements of edges are placed inside the border space without vertex-edge intersection. Then, we can link adjacent rectangles and support interactive queries of paths between vertices that are far away. The rectangle adjustment technique is designed carefully without introducing additional areal distortion. 
Through trade-off against areal error and aspect ratio loss, we use simulated annealing based heuristic optimization to minimize the topological error and make rectangles of connected vertices as adjacent as possible. 
We give details for each module of the framework in the following sections. 

\textbf{Error and loss metrics:} 
  \textit{Areal error:} 
  \revision{
  Areas of rectangles encode vertices' quantitative attributes. Areal error measures the degree of that rectangles' areal proportions deviate from vertices' weight proportions. A low areal error is important for the perceptual judgment of values attached on vertices. Following the definition in \cite{carrizosa2018mathematical}, to measure the areal error, we calculate the sum of absolute errors between weight proportions of vertices and area proportions of corresponding rectangles. Hence, the areal error is defined by}
  \begin{equation} \label{equa:error_area}
    \textrm{error}_{\textrm{a}} = \sum_{i=1}^n \abs{\alpha^P_i - \alpha_i}. 
  \end{equation}

\revision{
  \textit{Topological error:}  
  We make rectangles of connected vertices as adjacent as possible. If two vertices have an edge but their corresponding rectangles are not adjacent, we call this edge is a \textit{lost edge}; the number of lost edges is given by $|E \backslash E^P|$. If two veritices have no edge but their rectangles are adjacent, there exists a \textit{fake edge}; the number of fake edges is given by $|E^P \backslash E|$. Hence, following the definition in \cite{heilmann2004recmap}, the topological error is defined by
  \begin{equation} \label{equa:error_topology}
    \textrm{error}_{\textrm{t}} = \frac{|E \backslash E^P| + |E^P \backslash E|}{|E \cup E^P|}. 
  \end{equation}
  }

After the insertion of the border and the insertion of bridges that relate rectangles of connected vertices (e.g., \autoref{fig:les-miserables-border-placeholder}d and \autoref{fig:netherlands}b), fake edges are removed, and the topological error needs to be adjusted accordingly. 
We amend the topological error to only include lost edges: 
  \begin{equation} \label{equa:error_topology_amended}
    \textrm{error}_{\textrm{t}}' = \frac{|E \backslash E^P|}{|E|}. 
  \end{equation}


\revision{
  \textit{Aspect ratio loss:}  
  Good aspect ratios of rectangles are important for the accurate perception of data \cite{kong2010perceptual}. Kong et al. \cite{kong2010perceptual} and Heer et. al. \cite{heer2010crowdsourcing} measured perceptual judgment of data values with respect to different aspect ratios of rectangles, and their results show that both squarish and extreme aspect ratios have significantly larger errors than $\frac{2}{3}$ and $\frac{3}{2}$. 
  Hence, in our paper, we consider aspect ratios are good if the average aspect ratios of rectangles are close to $\frac{2}{3}$ and $\frac{3}{2}$. Note that the \textit{aspect ratio} of a rectangle in this paper hereafter is defined by the larger side of the rectangle divided by the smaller side of the rectangle, following the same definition in \cite{bederson2002ordered}. 
  Hence, the aspect ratios $\frac{2}{3}$ and $\frac{3}{2}$ belong to the same set (i.e., $\frac{3}{2}$) under this definition. The studies \cite{kong2010perceptual, heer2010crowdsourcing} were performed on five (i.e., a limited number of) aspect ratios, and there may exist a better aspect ratio for human perception. Hence, in our method, we allow one to specify the desired aspect ratio $r$, which leaves flexibility for new findings in the future. 
  Given a desired aspect ratio $r$, \textit{aspect ratio loss} 
  is the mean absolute error defined by}
  \begin{equation} \label{equa:loss_aspect_ratio}
    \textrm{loss}_{\textrm{r}} = \frac{1}{n}\sum_{i=1}^n \abs{max(\frac{w_i}{h_i}, \frac{h_i}{w_i})-r}. 
  \end{equation}



\begin{figure*}[tb]
\centering
\captionsetup[subfigure]{labelformat=empty}

\begin{subfigure}[b]{0.3\linewidth}
 \includegraphics[width=\columnwidth]{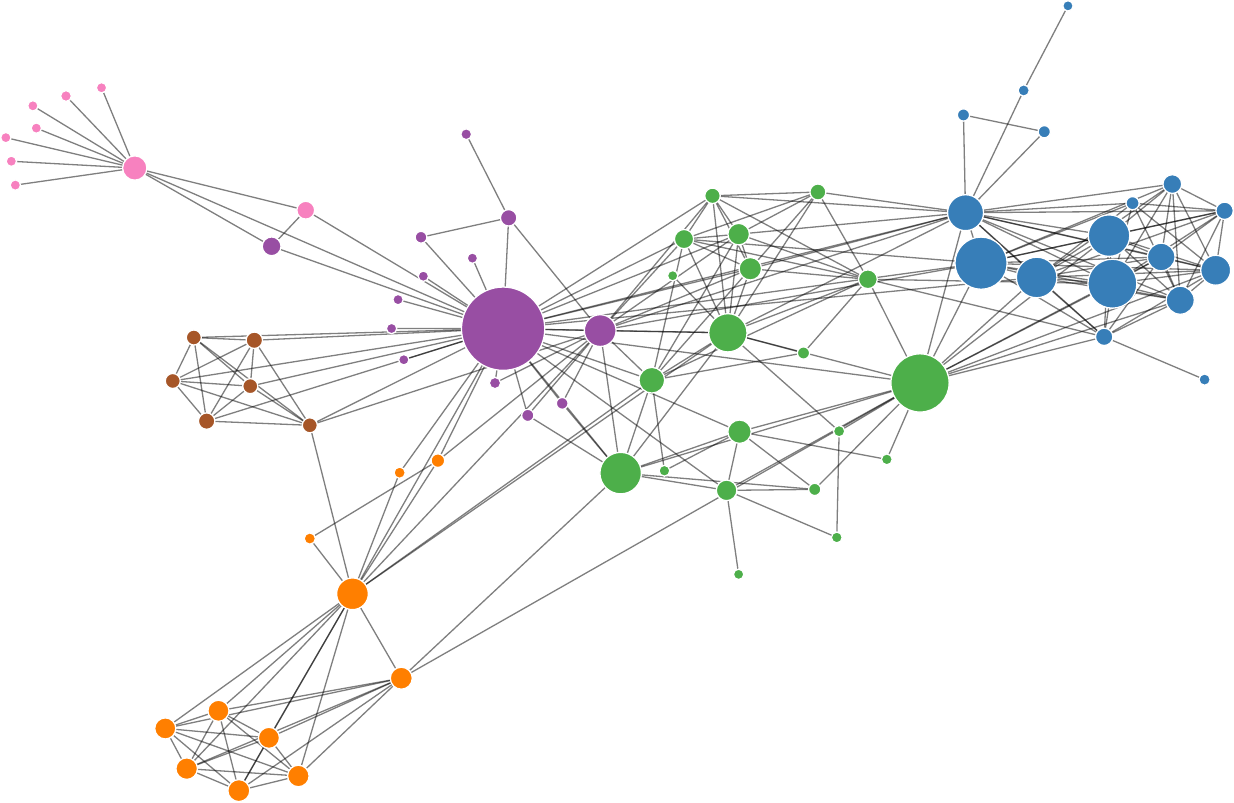}
 \caption{(a)}
\end{subfigure}
\hfill
\begin{subfigure}[b]{0.3\linewidth}
   \includegraphics[width=\columnwidth]{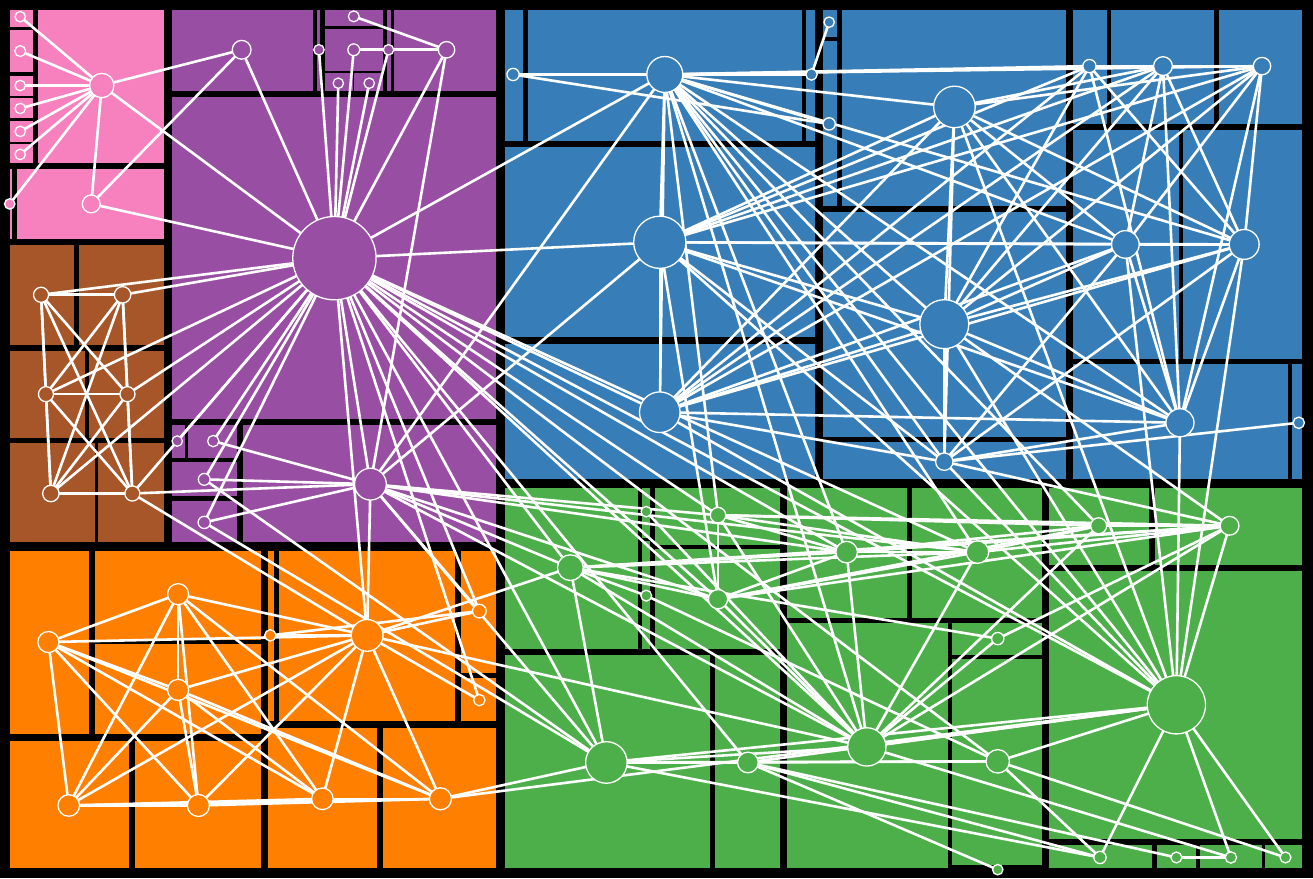}
   \caption{(b)}
 \end{subfigure}
 \hfill
 \begin{subfigure}[b]{0.3\linewidth}
   \includegraphics[width=\columnwidth]{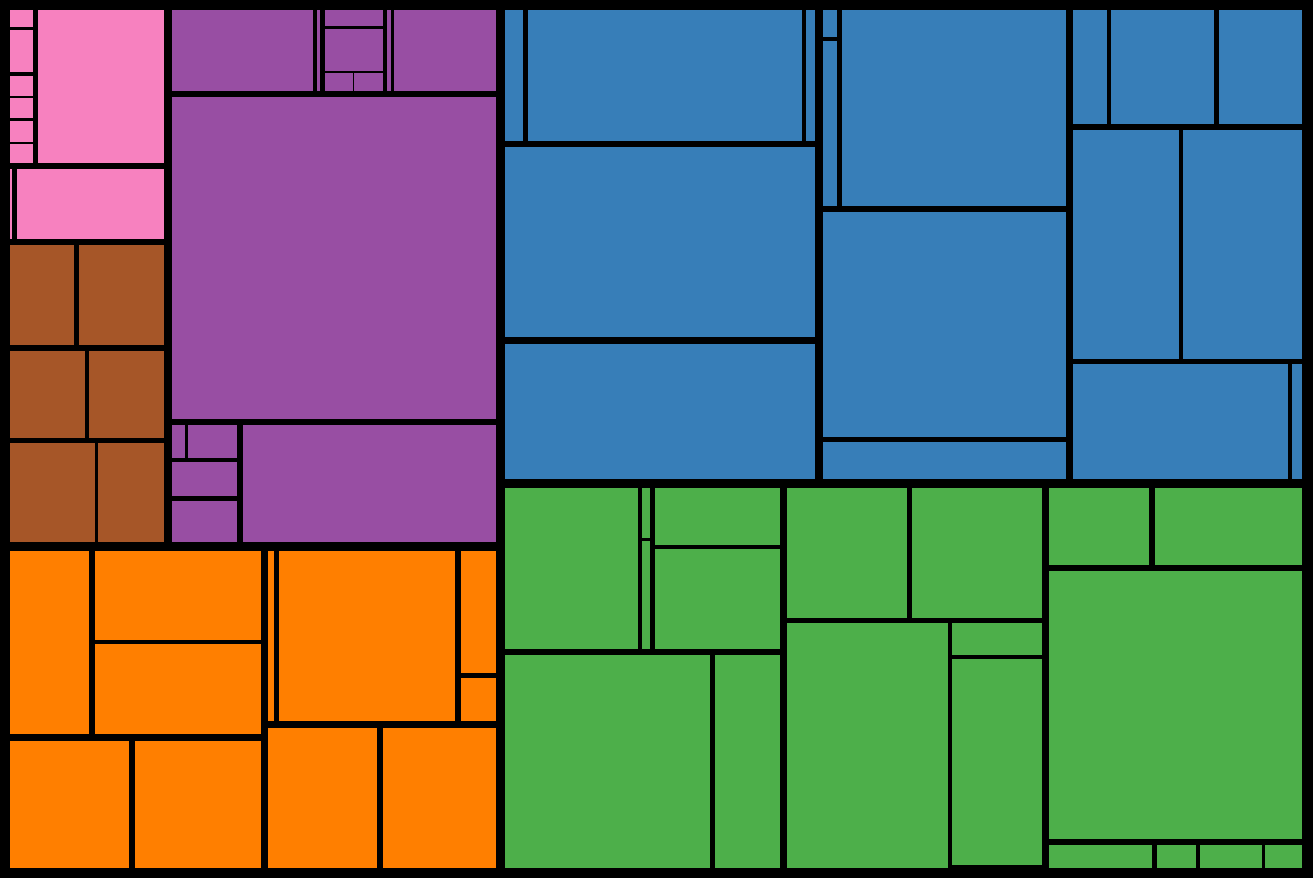}
   \caption{(c)}
 \end{subfigure}



 
  \caption{An example of a vertex-weighted graph for characters in a novel \textit{Les Mis\'{e}rables} \cite{hugo1887miserables}. 
  (a) A 2D graph embedding. 
  Each vertex is represented by a circle, and its weight is encoded by the circle's radius. Related vertices are connected by links. Colors encode clusters of vertices, and are suggested by ColorBrewer \cite{brewer2019colorbrewer}. 
  (b) The rectangular subdivision of the graph using the DAR rectangular partitioning method, overlaid by a node-link diagram. 
  (c) The partitioned rectangles only. The thickness of frame lines encodes the sequential order of the partitioning; thicker cuts occur earlier using the DAR partitioning algorithm. 
  }
  \label{fig:les-miserables-subdividing}
\end{figure*}



\section{Desired-Aspect-Ratio Rectangular Partitioning}
\label{sect:partitioning}

We provide a rectangular partitioning algorithm given the input of a two-dimensional embedding of a graph. 

\textbf{Discussions and limitations of alternative designs: }
Two categories of techniques: (1) insertion-based~\cite{wood2008spatially} and (2) partitioning-based methods~\cite{heilmann2004recmap, mansmann2007visual, duarte2014nmap, ghoniem2015weighted}, have been introduced to generate rectangular space-filling layouts for data elements with a 2D spatial order in the literature, although those methods have not been applied on graphs. 
First, spatially ordered treemap~\cite{wood2008spatially} is an insertion-based method that keeps inserting data elements with minimal spatial deviations from the inserted elements. 
Second, partitioning-based methods partition the 2D space recursively to acquire rectangular subdivisions for data elements. 
Previous partitioning-based methods apply two heuristic strategies: (1) equal-weight strategy~\cite{heilmann2004recmap, duarte2014nmap, ghoniem2015weighted} and (2) equal-number strategy~\cite{mansmann2007visual, duarte2014nmap}, to generate squarish rectangles.  
The equal-weight strategy seeks to subdivide a rectangle with 2D data elements into two smaller rectangles for each subdivision and ensure the weight sums of the data elements on the two subdividing rectangles are similar, similar to the pivot-by-split-size treemap~\cite{bederson2002ordered}. 
The equal-number strategy seeks to balance the number of 2D points on the two subdividing rectangles for each subdivision, similar to the pivot-by-middle treemap~\cite{shneiderman2001ordered, bederson2002ordered}. 
The abovementioned alternative designs have two limitations for our problem. 
First, because these methods are not designed specifically for graphs, the topological error is not considered. 
Second, previous partitioning-based methods \cite{mansmann2007visual, duarte2014nmap, ghoniem2015weighted} aim to provide squarish rectangles. However, the perception studies~\cite{kong2010perceptual, heer2010crowdsourcing} show that both squarish and extreme aspect ratios are not effective in data perception, hence, they advocate layout methods that can be optimized for a non-extreme aspect ratio.

\begin{figure}[htb]
\centering
\captionsetup[subfigure]{labelformat=empty}

 \begin{subfigure}[b]{0.3\columnwidth}
   \includegraphics[width=\columnwidth]{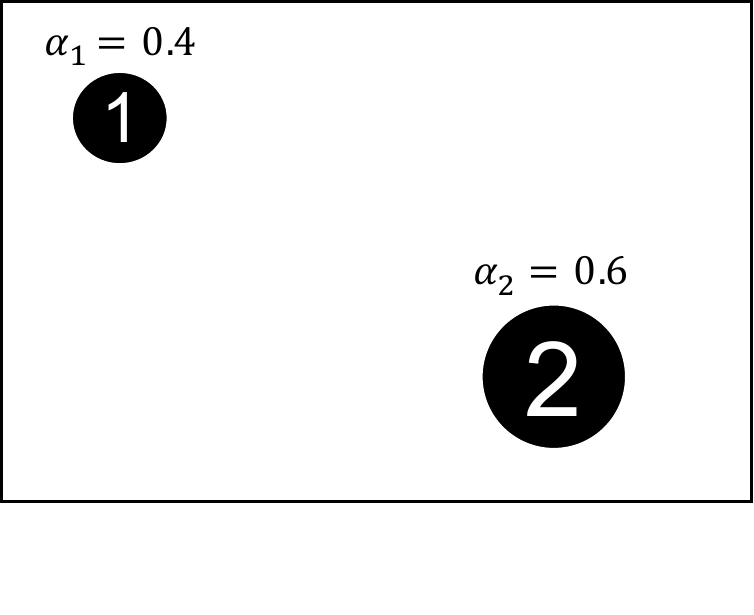}
   \caption{(a) initial} 
 \end{subfigure}
 \hfill
 \begin{subfigure}[b]{0.3\columnwidth}
   \includegraphics[width=\columnwidth]{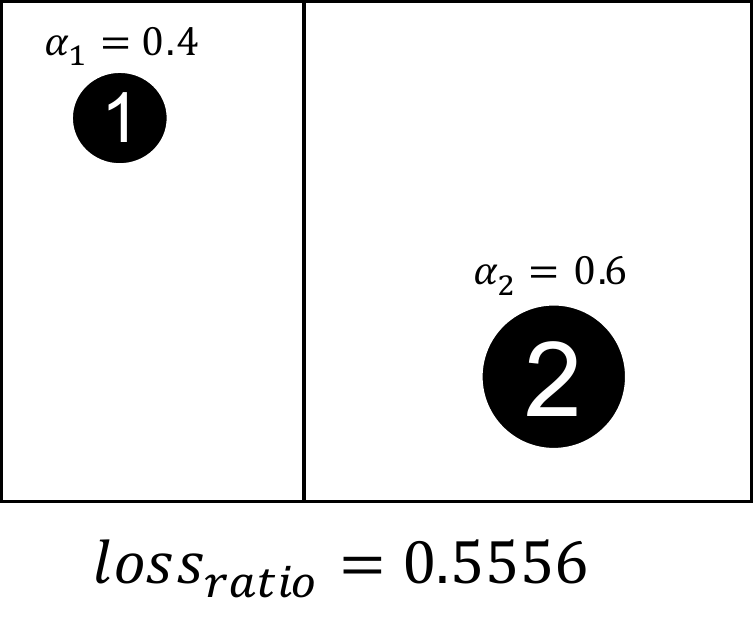}
   \caption{(b) horizontal} 
 \end{subfigure} 
 \hfill
 \begin{subfigure}[b]{0.3\columnwidth}
   \includegraphics[width=\columnwidth]{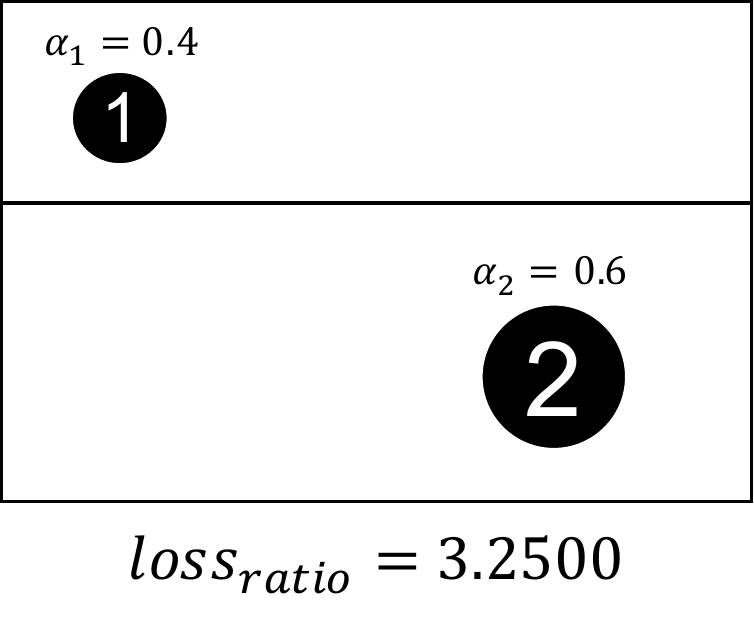}
   \caption{(c) vertical} 
 \end{subfigure} 

  \caption{Illustration of how to pick either horizontal partitioning or vertical partitioning by an example of two vertices. (a) Two vertices, vertex one with a weight of $0.4$ and vertex two with a weight of $0.6$, are placed on a rectangle with a width of three units and a height of two units. (b) We partition the rectangle horizontally and assign the two vertices to the two subdividing rectangles. The aspect ratio loss of the two rectangular subdivisions is $0.5556$. (c) We split the rectangle vertically to get two rectangular parts with an aspect ratio loss of $3.25$. Hence, we select the horizontal partitioning that has a lower loss. }
  \label{fig:partitioning-illustration}
\end{figure}

\textbf{K-d tree based data structure:}
K-dimensional tree (\textit{k-d tree}) is the output of our algorithm and is a well-known data structure for binary space partitioning. In a resulting k-d tree of our method, high-level nodes record rectangular space partitioning for clusters, and low-level nodes record splitting for vertices within clusters. 
Each \textit{leaf node} of the k-d tree corresponds to a vertex of the input graph and records the information of a subdividing rectangle for this vertex. 
Each \textit{internal node (non-leaf node)} of the k-d tree corresponds to a partitioning and records the information of (1) a \textit{splitting line} for this partitioning and (2) a subdividing rectangle enclosing the rectangles of all descendants of this internal node. 
The resulting k-d trees support efficient drawing, adjusting, and querying the rectangular subdivision of input graphs. 

\textbf{Our algorithm:} 
Our method is based on binary space partitioning (BSP) with a DAR heuristic strategy. BSP refers to an approach to subdivide space into convex sets using hyperplanes recursively, and specifically, we can use BSP to subdivide a 2D embedding space orthogonally into rectangles. 
By assuming the whole display space is rectangular, we first subdivide the rectangle of the display window to create rectangles for clusters based on clusters' 2D embeddings, where the embedding of a cluster is the average of all embeddings of vertices within the cluster. Second, we subdivide the rectangle of each cluster to construct rectangles for vertices within the cluster. 
This algorithm is based on a DAR strategy, which seeks to approach a specified aspect ratio for each rectangle by minimizing the loss of aspect ratio. 

We describe our algorithm in detail below. The input to our algorithm consists of (1) a rectangle with known width and height and (2) a list of vertices with weights and 2D embedding vectors. The output is a list of subdividing rectangles, each corresponding to a vertex, described by a k-d tree. 
We partition the input rectangle recursively. For each partitioning, given a rectangle $R$ and its associated list of $n$ vertices, we make either a horizontal or a vertical partitioning to produce two subdividing rectangles, $R_1$ and $R_2$, and distribute the vertices to the two rectangles. We give an example of two vertices in \autoref{fig:partitioning-illustration}; for more vertices, we exhaust all horizontal and vertical partitioning choices, and pick the best one with the lowest aspect ratio loss for $R_1$ and $R_2$, detailed as follows. 


\textbf{Horizontal partitioning:} 
We cut $R$ horizontally to obtain $R_1$ and $R_2$, and distribute vertices. $R_1$ and $R_2$ have the same height of $R$, and their widths are determined by their assigned vertices. We sort the $n$ vertices on $R$ by the $x$ coordinate ascendingly to get an ordered list $(v_1, ..., v_{n})$, denoting their associated weights by $\alpha_1, ..., \alpha_n$. Given $n$ vertices, we have $n-1$ choices to cut the ordered list into two nonempty sub-lists. We examine all choices and select the best one by evaluating the aspect ratio loss. Without loss of generality, we partition the list at location $k$ and get two sub-lists: $(v_1, ..., v_{k})$ and $(v_{k+1}, ..., v_{n})$. Hence, the weight sums of the two lists can be obtained by $\sum_{i=1}^{k}\alpha_i$ and $\sum_{i=k+1}^{n}\alpha_i$, respectively. We assign the first list $(v_i)_{i=1}^{k}$ to $R_1$ and the second list $(v_j)_{j=k+1}^{n}$ to $R_2$. To ensure zero areal error, given the two weight sums, we split the width of $R$ for $R_1$ and $R_2$ by the proportion of the two weight sums. To evaluate the quality of partitioning at location $k$, we compute the average aspect ratio loss of $R_1$ and $R_2$ by \autoref{equa:loss_aspect_ratio}. Then, we select the partitioning location with the smallest aspect ratio loss for the horizontal partitioning. 

\textbf{Vertical partitioning:} 
Similar to horizontal partitioning, when performing vertical partitioning to $R$, the resulting $R_1$ and $R_2$ have the same width of $R$, and their heights are determined by their assigned vertices. We sort the $n$ vertices by their $y$ coordinates ascendingly to get an ordered list, and examine where to partition the ordered list based on evaluating the aspect ratio loss. Then, we choose the location with the smallest aspect ratio loss. 

Finally, to decide whether to perform horizontal or vertical partitioning for a given subdivision, we compare the aspect ratio losses for the two types of partitioning and select the one with a smaller loss. The result generated from \textit{Les Mis\'{e}rables} network is shown in \autoref{fig:les-miserables-subdividing}.

\section{Two-Stage Rectangle Adjustment}

\begin{figure}[htb]
\centering
 \begin{subfigure}[b]{0.45\columnwidth}
   \includegraphics[width=\columnwidth]{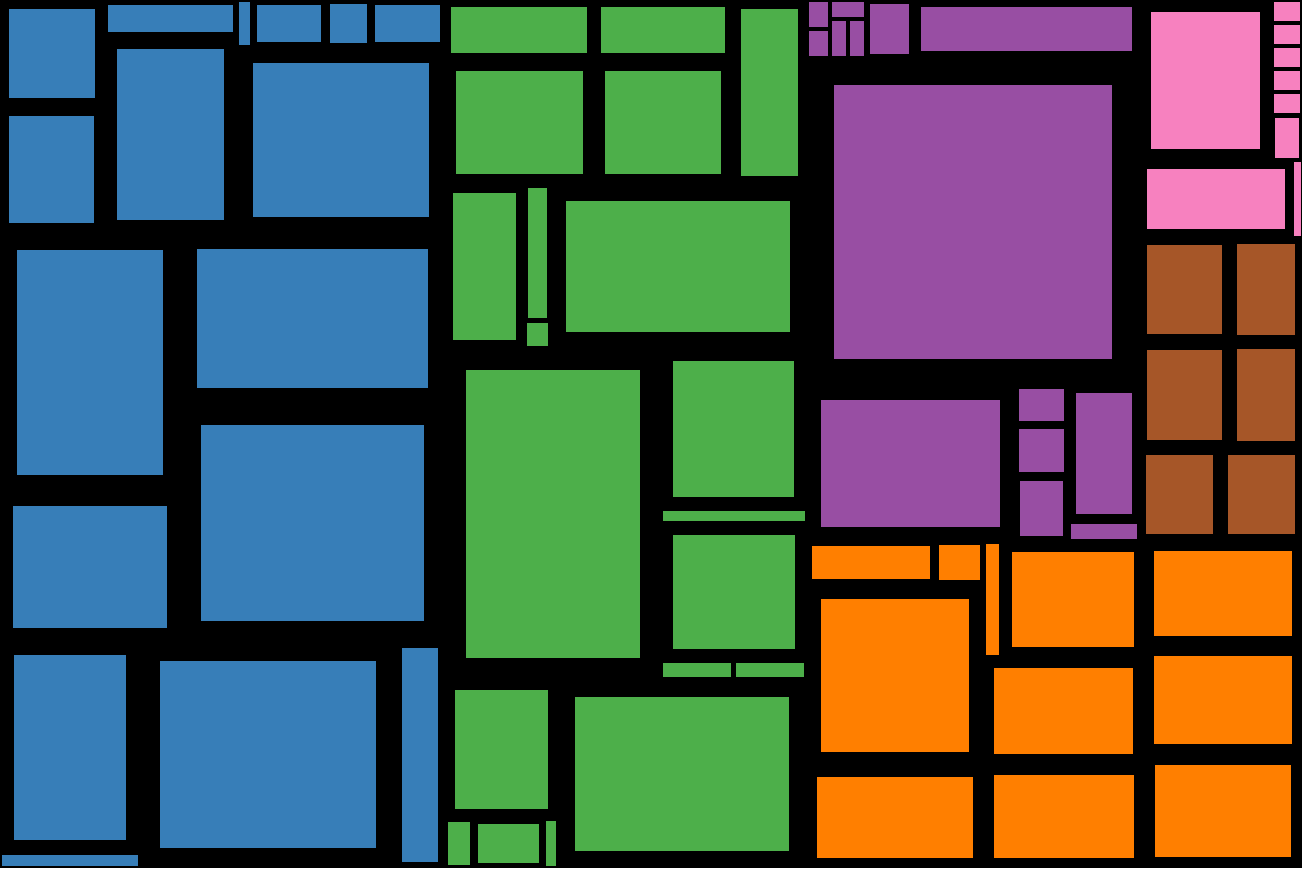}
   \caption{} 
   \label{fig:les-miserables-border-placeholder-proportional}
 \end{subfigure}
 \hfill
 \begin{subfigure}[b]{0.45\columnwidth}
   \includegraphics[width=\columnwidth]{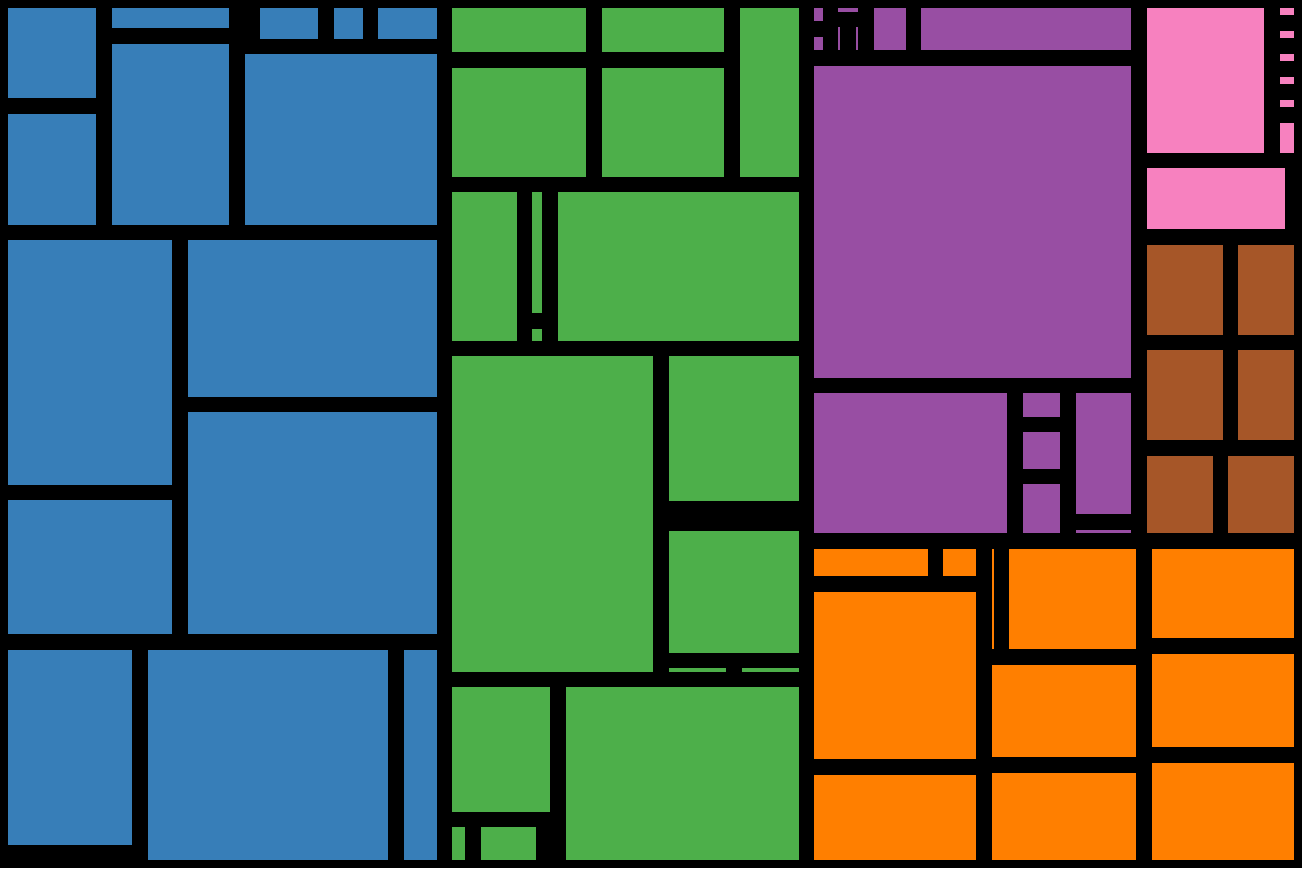}
   \caption{} 
   \label{fig:les-miserables-border-placeholder-trivial}
 \end{subfigure} 

 \bigskip

 \begin{subfigure}[b]{0.45\columnwidth}
   \includegraphics[width=\columnwidth]{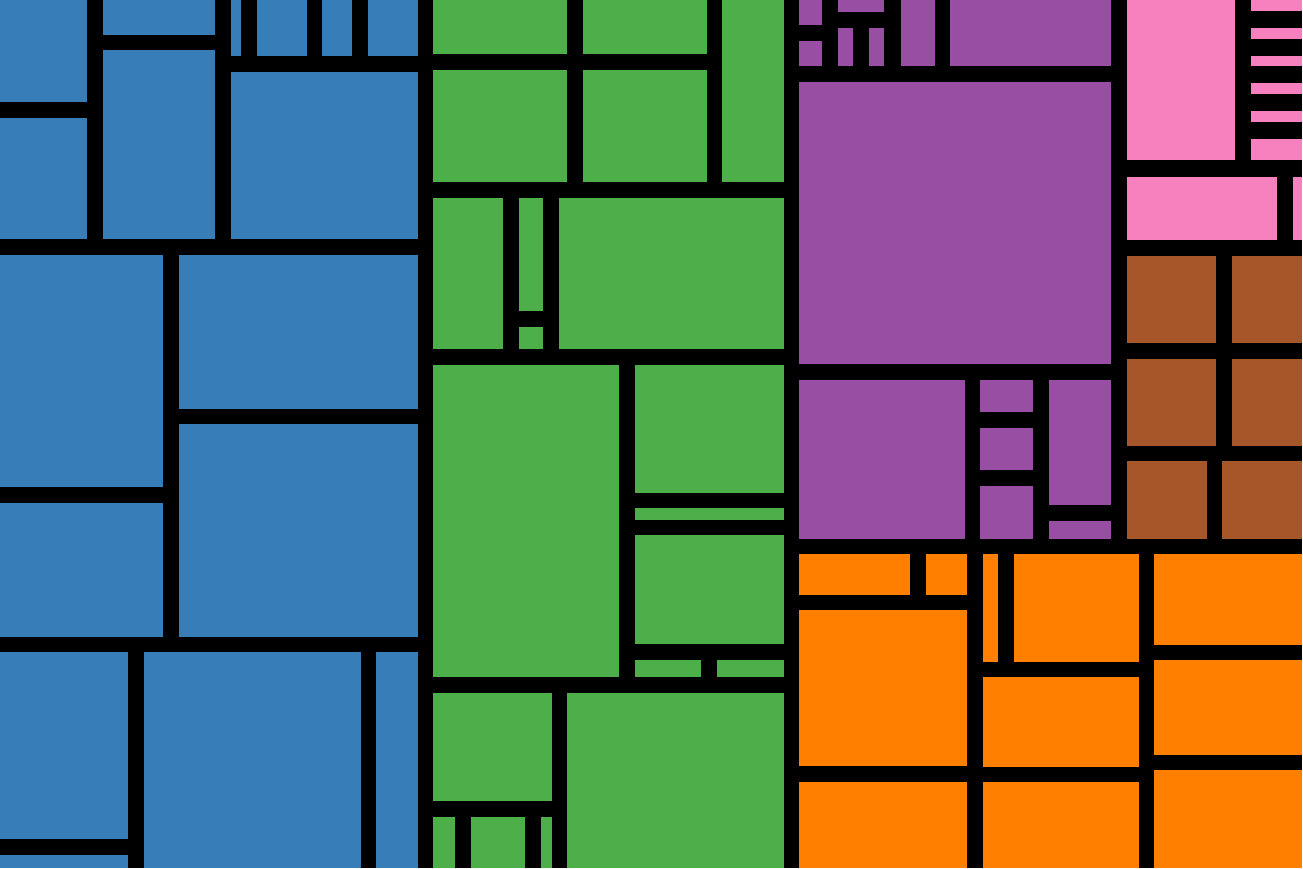}
   \caption{} 
   \label{fig:les-miserables-border-placeholder-two-stage}
 \end{subfigure} 
 \hfill
 \begin{subfigure}[b]{0.45\columnwidth}
   \includegraphics[width=\columnwidth]{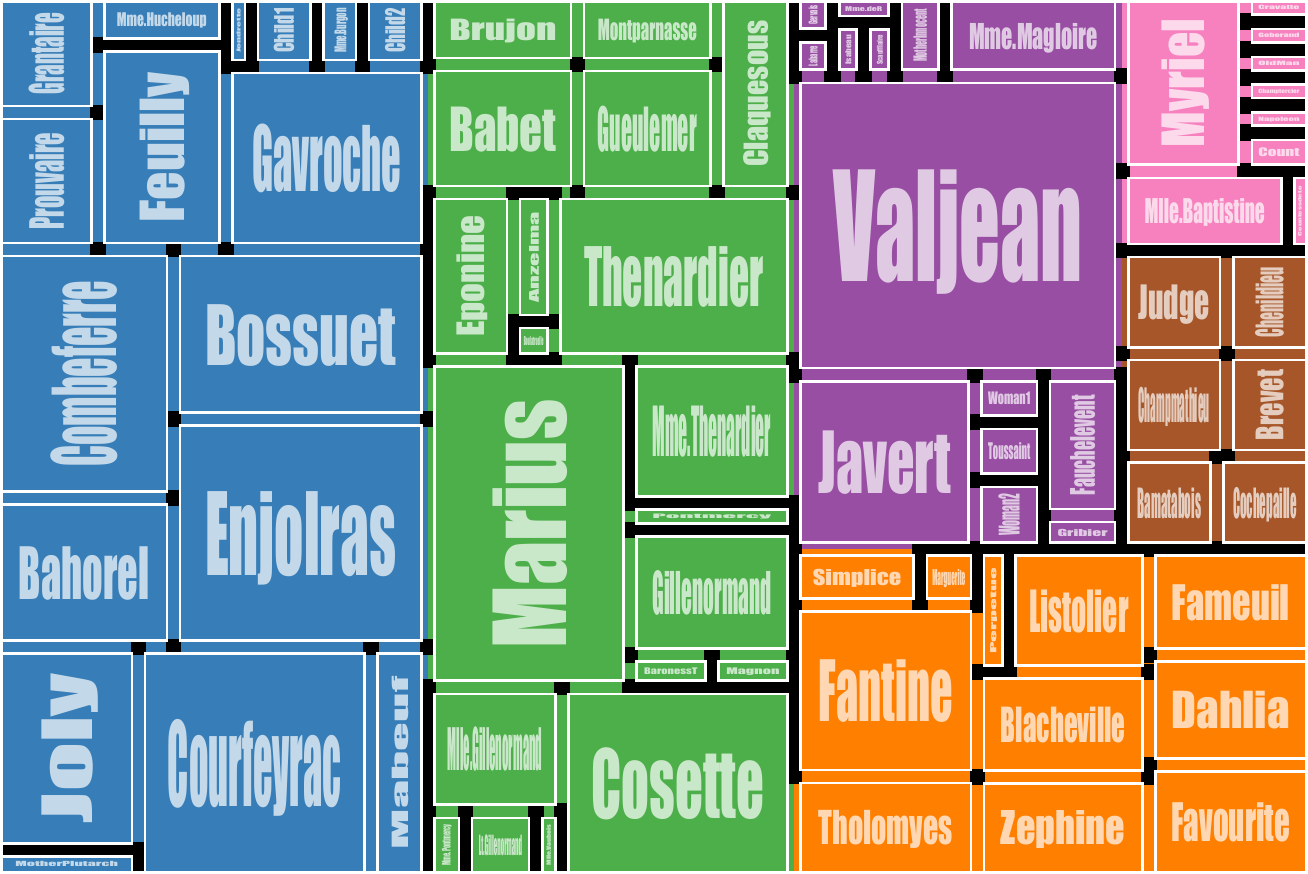}
   \caption{}
 \end{subfigure}

  \caption{Illustration for border insertion: (a) proportional border, (b) fixed-width border by a trivial method, and (c) fixed-width border by our two-stage adjustment technique. Borders are painted black. 
  Based on (c), (d) direct connections between adjacent rectangles are represented by bridging rectangles, eliminating fake edges. }
  \label{fig:les-miserables-border-placeholder}
\end{figure}

\begin{figure*}[tb]
  \centering 
 \begin{subfigure}[b]{0.3\linewidth}
   \includegraphics[width=\columnwidth]{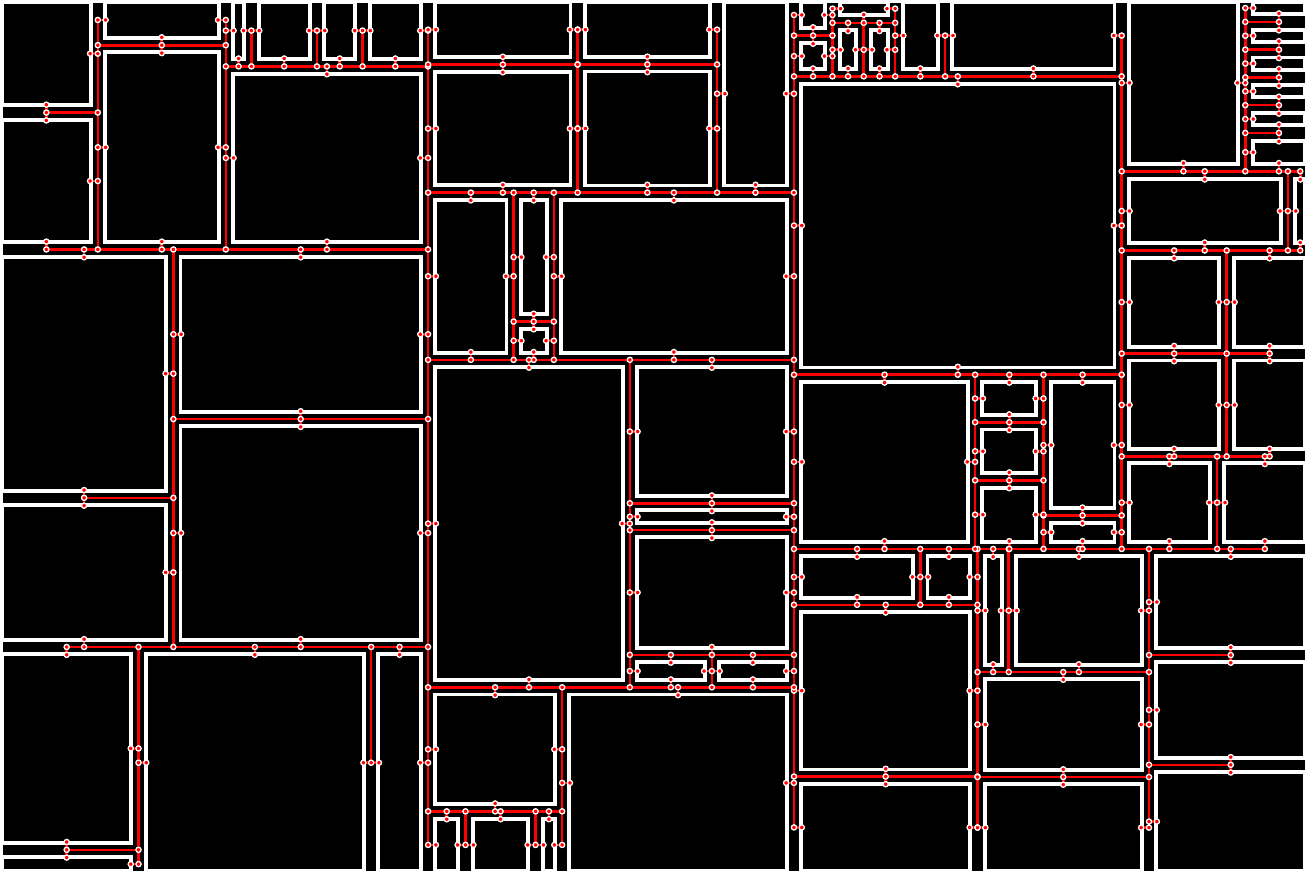}
   \caption{}
 \end{subfigure}
 \hfill
 \begin{subfigure}[b]{0.3\linewidth}
   \includegraphics[width=\columnwidth]{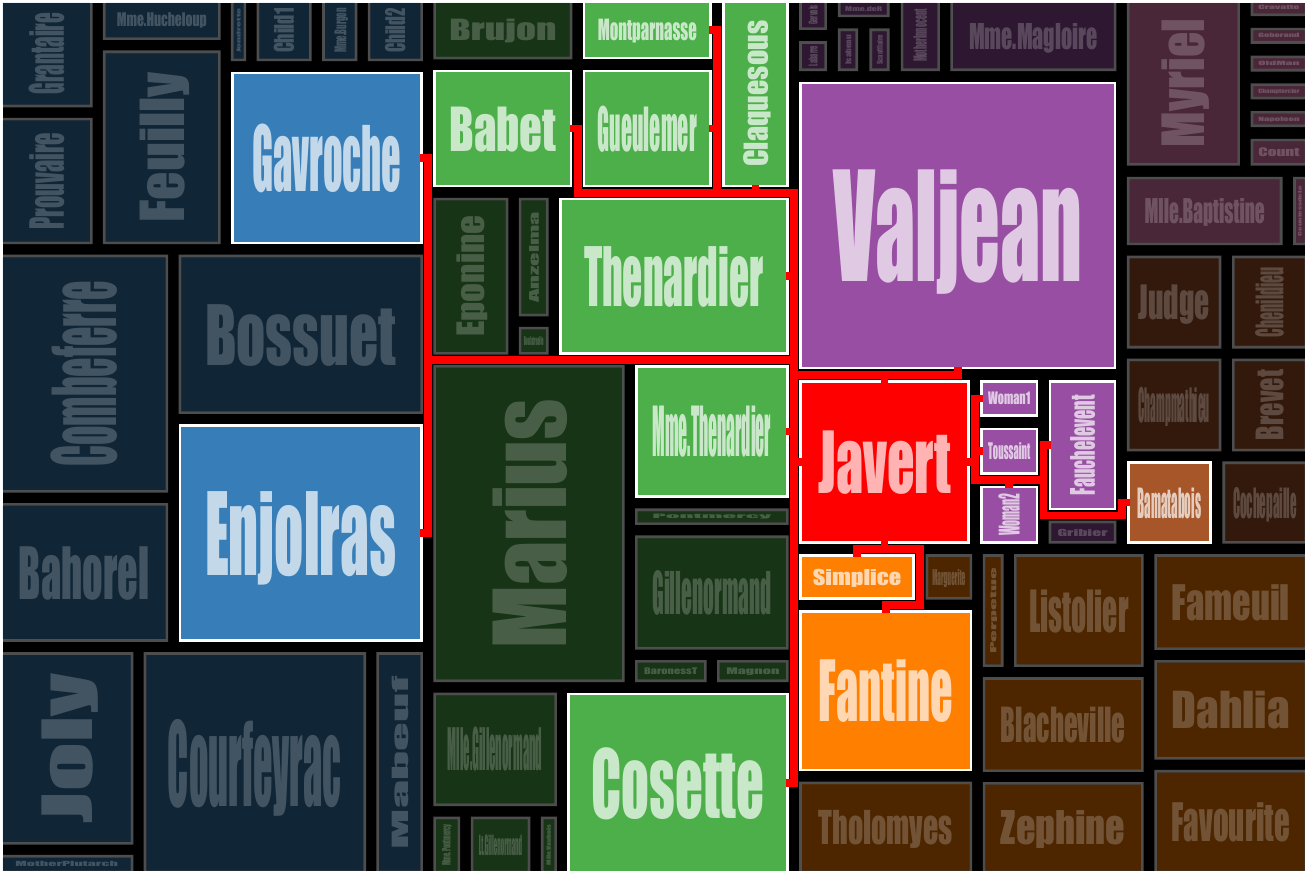}
   \caption{}
 \end{subfigure} 
 \hfill
 \begin{subfigure}[b]{0.3\linewidth}
   \includegraphics[width=\columnwidth]{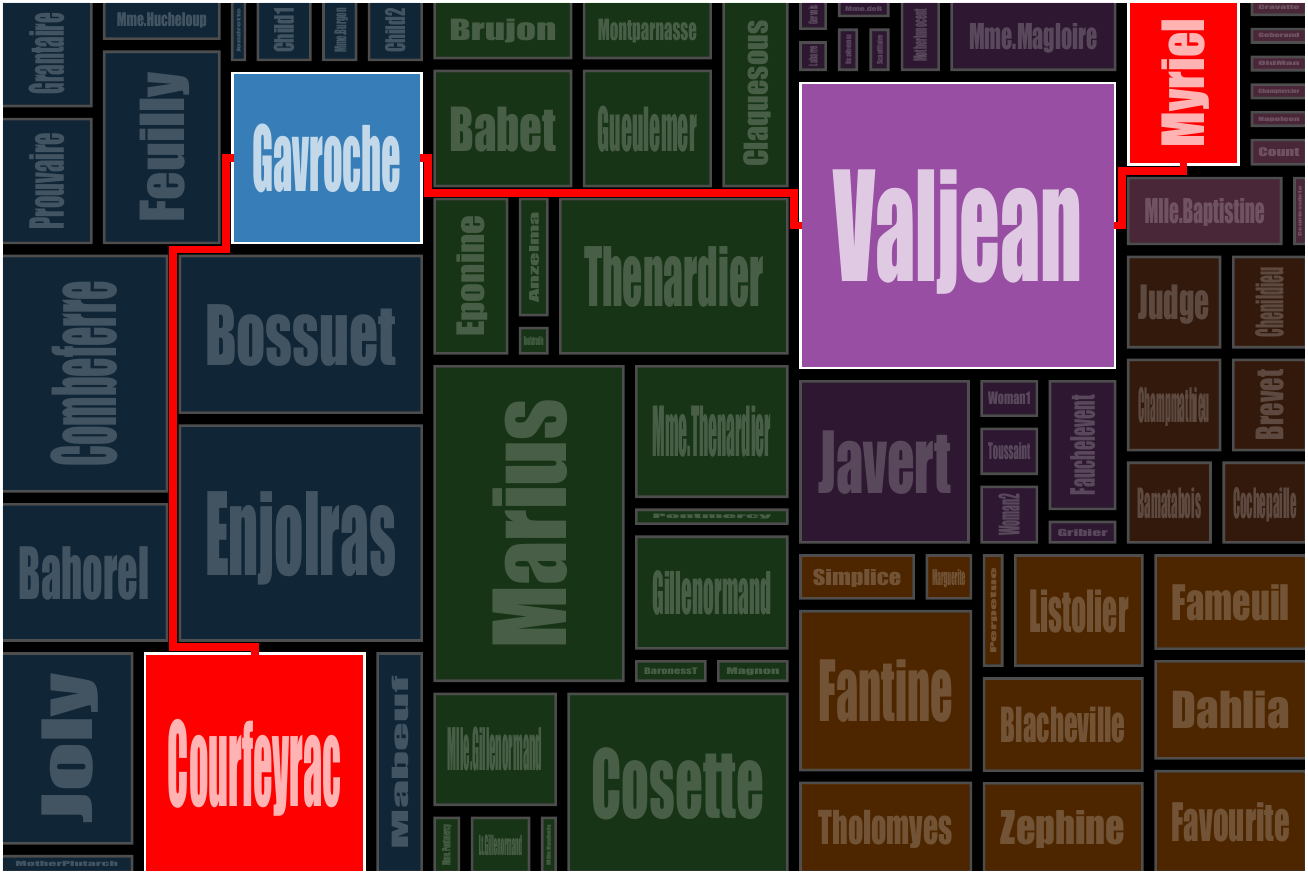}
   \caption{}
 \end{subfigure} 

  \caption{Illustration for routing paths between rectangles of connected vertices on borders when vertices are far away. 
  (a) The red points include center points of rectangle sides, center points of sides of rectangle borders, and intersection points of rectangle borders; adjacent red points are connected by red lines to become a network. 
  (b) Over the network of (a), we place the paths from a hovered rectangle ``Javert'' to its connected rectangles. 
  (c) After we select two rectangles, the path between them with the shortest-hop connection is displayed over the network of (a). 
  }
  \label{fig:grid}
\end{figure*}

We create border space surrounding rectangles as tunnels to place graph edges. 
In the border space, we can relate rectangles by link geometries and eliminate fake edges. 
Specifically, connections between adjacent rectangles can be encoded using bridging rectangles (e.g., 
\autoref{fig:les-miserables-border-placeholder}d). 
We can route paths between rectangles inside the border space to avoid vertex-edge occlusion to observe the egocentric network by specifying a rectangle of interest (e.g., \autoref{fig:grid}b) or identify a multi-hop path between two selected rectangles (e.g., \autoref{fig:grid}c).




Before giving detailed algorithms, we discuss two types of rectangle borders: (1) proportional borders and (2) borders with a fixed as the motivation of our algorithm. 
Note that to preserve areal proportions $\{\alpha^P_i\}_{i=1}^{n}$ of the given partitioning $G^P$ and protect small rectangles from being removed, the area of borders needs to be taken into account explicitly for visualization generation; otherwise, as pointed out by L{\"u} et al. \cite{lu2008cascaded} for treemaps \cite{johnson1991tree}, the given area proportions are distorted after borders of rectangles are inserted for additional information. The distortion becomes worse when borders become wider. To tackle the problem, we describe how to adjust the subdividing rectangles to insert the two types of borders in the following. 

\textbf{Proportional border: }
Given a rectangular partitioning, each rectangle contributes a proportion $p$ of its area to build its border. After that, each rectangle owns $1-p$ of its original area, which preserves the original areal proportions; also, small rectangles are guaranteed not to be removed. 
The result is shown in \autoref{fig:les-miserables-border-placeholder-proportional}. Although we can compute proportional borders straightforwardly, the border widths of resulting rectangles are not uniform, leading to bad readability. 

\textbf{Border with a fixed width:} 
Given a rectangular subdivision, we open up space for borders with a fixed width $d$ and ensure that the areal proportions $\{\alpha^P_i\}_{i=1}^{n}$ remain unchanged after the insertion of borders, which is non-trivial. Consider a trivial method that, similar to the proportional border, each rectangle $R$ directly contributes a hollow rectangle with thickness $d$ as its border, for example, \autoref{fig:les-miserables-border-placeholder-trivial}. As a result, small and narrow rectangles may vanish. Furthermore, the area of the border is $2d \cdot (w_R+h_R)-4d^2$, and the area of the rectangle left for areal encoding is $w_R \cdot h_R$ deducting by the border area; the encoding area is not proportional to $w_R \cdot h_R$, hence, introduces more areal errors. 

To solve the problem, we design a two-stage adjustment algorithm including (1) a coarse top-down adjustment and (2) a fine bottom-up adjustment. In the literature, to reduce the areal distortion of treemaps given by borders, the cascaded treemap \cite{lu2008cascaded} offers a high-level description of a top-down adjustment, which adjusts the sizes of subdividing rectangles on treemaps from the top to the bottom of the input hierarchical tree by considering sizes of borders. However, the adjustment method in \cite{lu2008cascaded} does not consider that the border area is not a static value and may change after adjusting rectangles hence introducing additional areal errors. Moreover, a top-down adjustment method solely is not sufficient to preserve the areal proportions $\{\alpha^P_i\}_{i=1}^{n}$. In our two-stage adjustment, we solve a system of linear equations to consider the change of border area when adjusting each subdivision. Furthermore, we offer the other bottom-up adjustment to ensure unchanged areal proportions. Hence, we can guarantee the areal proportions of the input rectangular partitioning are preserved. 

\subsection{Two-Stage Adjustment Details}
In the following, we design a two-stage adjustment to ensure unchanged areal proportions when rectangle borders with a fixed width are inserted. Formal proof for the proportion preservation is provided in the appendix. 
The input of our adjustment is a rectangular subdivision $G^P$ represented by a k-d tree described in the last section. The output is a rectangular visualization with faithful areal encodings and rectangle borders with a fixed width. A result example of our two-stage adjustment is shown in \autoref{fig:les-miserables-border-placeholder-two-stage}.

\textbf{Coarse top-down adjustment:} 
We traverse the k-d tree of the input rectangular partitioning from the root to the leaves to scale subdividing rectangles for two purposes: (1) inserting a border with a fixed width $d$ for each leaf subdividing rectangle and (2) adjusting areas of subdividing rectangles based on their given areal proportions. 
We explain the intuitions of the top-down adjustment hereafter and give detailed computation derivations in the appendix. 
Intuitively, the insertion of borders with a fixed width of $d$ is equivalent to expanding the splitting lines between rectangles to splitting bands with widths of $2d$, however, with certain adjustments to guarantee the areal proportions unchanged. 
Note that each internal node in our k-d tree is associated with a rectangle $R$ and a splitting line $l$ partitioning $R$ into two subdividing rectangles $R_1$ and $R_2$. After expanding the splitting line $l$ into a band, the encoding areas of rectangles change; the \textit{encoding area} of a rectangle $R$ for areal encoding is equal to the area of the rectangle $R$ reducing the total border area inside $R$. 
To ensure that the encoding areas of the two subdividing rectangles $R_1$ and $R_2$ are proportional to their original given areal proportions, we adjust the position of the splitting line $l$ to a new position by solving a system of linear equations detailed in the appendix. 

However, the top-down adjustment solely is not sufficient because the adjustment of low-level partitionings of the k-d tree hampers high-level partitionings that have been adjusted. An example is given in the appendix. 
Therefore, we introduce the fine bottom-up adjustment afterwards to solve the problem. 

\textbf{Fine bottom-up adjustment:} 
The bottom-up adjustment traverses the k-d tree from leaves to the root to scale subdividing rectangles to guarantee the given input areal proportions are preserved. 
Given an interval node with a rectangle $R$ and a splitting line $l$, we move $l$ to make the encoding areas of leaf subdividing rectangles have the same proportions as the given areal proportions. 
After the descendants of the internal node have been adjusted correctly, since it is bottom-up, now we adjust $l$ by solving a system of linear equations, detailed in the appendix, to determine a new position of $l$.

\subsection{Edge Encodings in Border Space}
After inserting rectangular borders as tunnels, we have enough space to route edges between vertices. 

\subsubsection{Bridges between Adjacent Rectangles}
We create bridging rectangles between adjacent rectangles that have edges in the raw graph $G$ to highlight their direct connections. If the rectangles of two connected vertices have contacts between their borders, we record the touching segments of the borders. The touching segments are bridges between related rectangles. We paint the segments by the colors of their corresponding rectangles; other segments of borders are colored black to block disconnected vertices so that the fake edges are removed. Since bridge segments are also rectangular, to distinguish rectangles of vertices against the bridges, we enclose rectangles of vertices with white lines. The results are shown in \autoref{fig:les-miserables-border-placeholder}d. 

Since each rectangle can only be adjacent to a limited number of surrounding rectangles, when the edges of the given graph become dense, the lost edges increase. Hence, we further complement more relationships between vertices to VMap by using interactions to query certain topological structures hereafter. 

\subsubsection{Egocentric Network}
When hovering a rectangle as an ego, we use red channels to relate rectangles with direct connections to the hovered rectangle. Given that each pair of rectangles can be connected through the border space, we can draw the red channels over the border space by orthogonal lines. 
As illustrated in \autoref{fig:grid}a, the red points include the centers of rectangle sides, centers of border sides, and intersection points of borders; the adjacent red points are connected to form a network visualized by the red lines. When hovering a rectangle, we use Dijkstra's algorithm \cite{dijkstra1959note} to acquire the shortest path from the hovered rectangle to its related rectangles over the red network in \autoref{fig:grid}a to form an egocentric network. For example, when we hover ``Javert'' in \autoref{fig:grid}b, the red channels to its related rectangles are placed in the border space. 







\subsubsection{Vertex-to-Vertex Query}
When querying the connection between two vertices, we display the path between the vertices with the shortest hop, which is important for several applications. For example, researchers of social networks are interested in the paths with the shortest social connections between two individuals; the ``six degrees of separation'' \cite{guare1990six} is an empirical idea that each pair of people can be connected by six or fewer social connections. Also, for wireless sensor networks, the path with the shortest hop can be used to produce energy-efficient routes for data dissemination \cite{yilmaz2012shortest}. 
We also utilize Dijkstra's algorithm~\cite{dijkstra1959note} to search the path with the shortest hop over the network of \autoref{fig:grid}a. A result is shown in \autoref{fig:grid}c, where the path is drawn by red, and the rectangles of the two given vertices are highlighted as well as the in-between rectangles. 


\section{Simulated Annealing based Heuristic Optimization}

\revision{
Inspired by simulated annealing~\cite{kirkpatrick1983optimization} based drawing and layout~\cite{davidson1996drawing, brank2004drawing, kopp2018temporal}, we tailor simulated annealing to minimize and balance areal error, topological error, and aspect ratio loss to achieve a low total cost. Because the error and loss functions are not differentiable, gradient-descent-based optimizations~\cite{brank2004drawing} are not utilized here. 
}

\textbf{Optimization problem statement:} 
Our goal is to find the optimal configuration with the minimal cost, where a \textit{configuration} $\sigma$ is a record that contains vertices' weights, embedding positions, and a desired aspect ratio. $\sigma$ is a candidate used to generate a corresponding rectangular layout. 
\begin{equation}
    \argmin_{\sigma} \textrm{cost}(\sigma), 
\end{equation}
Given a rectangular partitioning generated based on a configuration, the cost of the configuration is defined by the weighted sum of \textbf{a}real error, \textbf{t}opological error, and aspect \textbf{r}atio loss: 
\begin{equation}
\textrm{cost}(\sigma) = \lambda_{\textrm{a}} \textrm{error}_{\textrm{a}} + \lambda_{\textrm{t}} \textrm{error}_{\textrm{t}}  + \lambda_{\textrm{r}} \textrm{loss}_{\textrm{r}}, 
\end{equation}
where $\lambda_{\textrm{a}}$, $\lambda_{\textrm{t}}$, and $\lambda_{\textrm{r}}$ are non-negative weights, and $\lambda_{\textrm{a}}+\lambda_{\textrm{t}}+\lambda_{\textrm{r}}=1$. 

We follow the previous work~\cite{davidson1996drawing, kopp2018temporal} to design the simulated annealing for the optimization problem below. 

\revision{
\textbf{Simulated annealing overview:} 
We perturb vertices' weights, 2D positions, and the desired aspect ratio to minimize the total cost and allow certain additional errors on unimportant aspects. 
At the beginning of the optimization, we choose an initial temperature $T$ and initialize the configuration $\sigma$ with vertices' original weights and random 2D positions. 
The optimization has $ns$ stages, and $T$ cools down slowly after every stage. Each stage has $ni$ iterations. 
In each iteration, a \textit{perturbation action} is randomly-picked from three options: (1) position perturbation, (2) weight perturbation, and (3) desired aspect ratio perturbation, and changes the current configuration $\sigma$ to a perturbed configuration $\sigma'$. 
If positions are perturbed, positions of all vertices in $\sigma'$ are normalized using min-max normalization so that the range of each dimension is within $[0, 1]$ and the entire embedding is in $[0, 1]\times[0, 1]$ to avoid degeneration. 
$\sigma$ will be replaced by $\sigma'$ under two circumstances: (1) if $\textrm{cost}(\sigma') < \textrm{cost}(\sigma)$, or (2) with a probability $e^{(\textrm{cost}(\sigma) - \textrm{cost}(\sigma'))/(T/256)}$, which is the probability of accepting a worse configuration and decreases as $T$ decays. 
Finally, $\sigma$ is fine-tuned with additional $ns$ stages and $T$ reinitialized to the initial value. Also, in fine-tuning, only configurations with lower costs are accepted. 
}

\revision{
\textbf{Cooling schedule:} 
Temperature $T$ steers the optimization process and determines the volatility of the optimization. Intuitively, higher $T$ causes perturbations with larger ranges and more frequent transitions between configurations to skip local minima of the cost function for exploration; on the contrary, lower $T$ tends to find a local minimum of the cost for exploitation. 
The temperature $T$ starts with a high initial value (i.e., an upper bound $T_{\textrm{ub}}$), and decays by multiplying with a factor $0< \gamma < 1$ at the end of every stage, until reaches a lower bound $T_{\textrm{lb}}$: 
\begin{equation}
T_0 = T_{\textrm{ub}}, T_s = \gamma T_{s-1}, \forall s \geq 1, 
\end{equation}
where $s$ is the stage index. 
Given the total number of stages $ns$, the decay factor $\gamma$ can be computed by: $\gamma = \sqrt[ns]{T_{\textrm{lb}} / T_{\textrm{ub}}}$.} 
In this paper, $T_{\textrm{ub}}$ is set to $256$, and $T_{\textrm{lb}}$ is associated with the vertices' minimal weight such that $T_{\textrm{lb}}=\sqrt{\min{\{\alpha_i\}} / r}/128$, where $r$ is the target aspect ratio specified by users. These settings perform well in our studies empirically. 

\revision{
\textbf{Position perturbation:} 
A random vertex, $v_i$, moves a step away from its current location with a fixed step size and a random direction angle among $360^{\circ}$ in the 2D space, and hence the next position is located on a circle centered on its current location with the radius being the step size. The step size is set to $\min{\{1, T\}}$, and hence is restricted to be not large than one because a large value may cause degeneration. }

\revision{
The movement direction is a combination of three heuristics: (1) random, (2) attraction, and (3) repulsion. 
First, the random heuristic is a vector with a magnitude being $T$ and direction being a random angle among $360^{\circ}$ in the 2D space. 
Second, the attraction heuristic is a vector with a magnitude being $1+T$. To compute the direction of the vector, we find out which other vertices have lost edges with $v_i$ and compute the sum of the directional unit vectors from $v_i$ to those vertices, where the direction is parallel with the sum. An exception is that if a vertex is already very close to $v_i$ with their distance lower than $\frac{\sqrt{\alpha_i}}{2}$, their attraction is not counted to avoid degeneration. 
Third, the repulsion heuristic is a vector with a magnitude being $T$. To compute the direction of the vector, we find out which other vertices have fake edges with $v_i$ and compute the sum of the directional unit vectors from those vertices to $v_i$, where the direction is in parallel with the sum. An exception is that if a vertex is already far away from $v_i$ with their distance larger than $\frac{\sqrt{2}}{2}$, their repulsion is not counted to avoid degeneration. 
After normalizing the weighted sum of the vectors of the three heuristics, we obtain a unit vector representing the direction of the movement. }

\revision{
\textbf{Weight perturbation:} 
The weight of a random vertex is multiplied with a factor $(1+T)$ for magnification or $\frac{1}{(1+T)}$ for contraction with $50\%$ possibility for each. We apply a value clipping to ensure the perturbed weight is within $[\frac{1}{64}, 64]$ times of the original weight. Note that, for applications having strict restrictions for faithful areal encodings, the configurations with zero areal error can be obtained if the weight perturbation is not performed. 
}

\revision{
\textbf{Desired aspect ratio perturbation:} We perturb the desired aspect ratio used for rectangular partitioning; note that, although the desired aspect ratio is perturbed, the aspect ratio loss is computed based on the initial user-determined target aspect ratio and not the perturbed one. 
Similar to the weight perturbation, the desired aspect ratio is multiplied with a factor $(1+T)$ for magnification or $\frac{1}{(1+T)}$ for contraction with $50\%$ possibility for each. After that, a value clipping is applied to ensure the ratio range is within $[1, 64]$. 
}

\begin{figure*}[tb]
  \centering 
  \includegraphics[width=\linewidth]{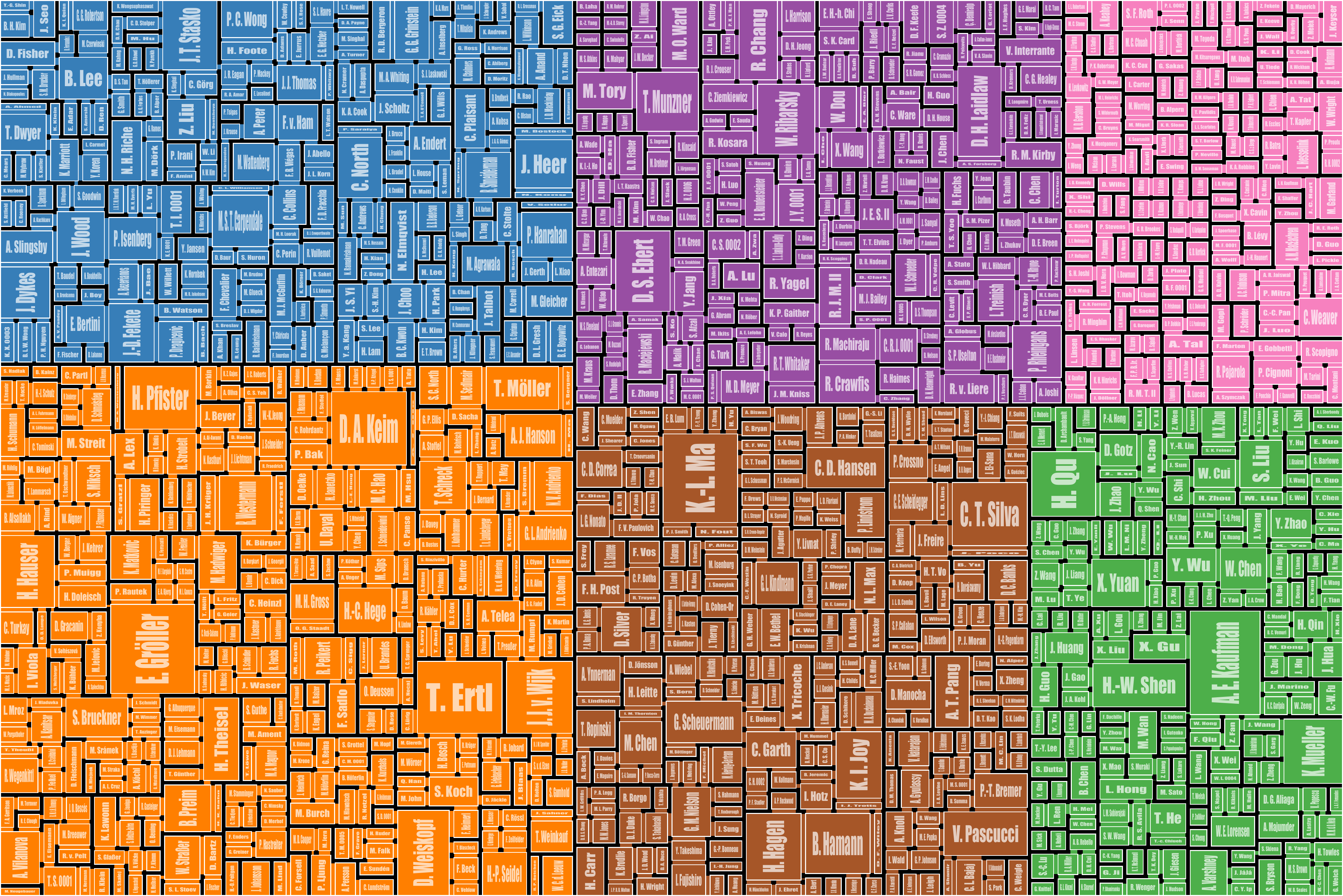} 
  \caption{VMap for the co-authorship of VIS publications before 2016. Each rectangle represents a researcher, and its size encodes the number of published co-authored papers; words on rectangles are the names of corresponding authors. Bridges between rectangles indicate co-authoring events between adjacent authors; black blocks obstruct ones with no edges. }
  \label{fig:teaser}
\end{figure*}
\begin{figure*}[tb]
  \centering 
\begin{subfigure}[b]{0.27\linewidth}
   \includegraphics[width=\columnwidth]{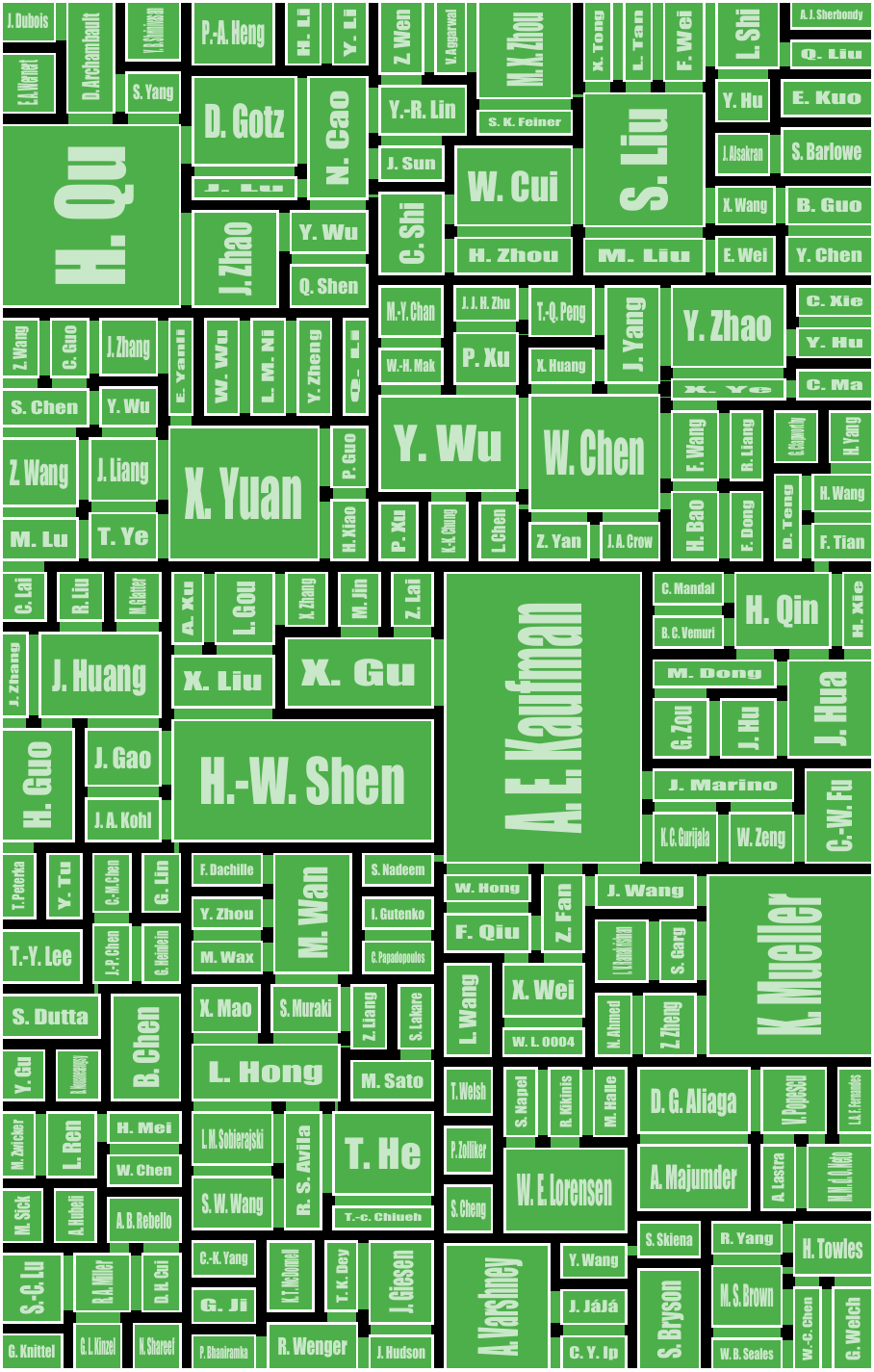}
   \caption{}
 \end{subfigure} 
 \hfill
 \begin{subfigure}[b]{0.67\linewidth}
   \includegraphics[width=\columnwidth]{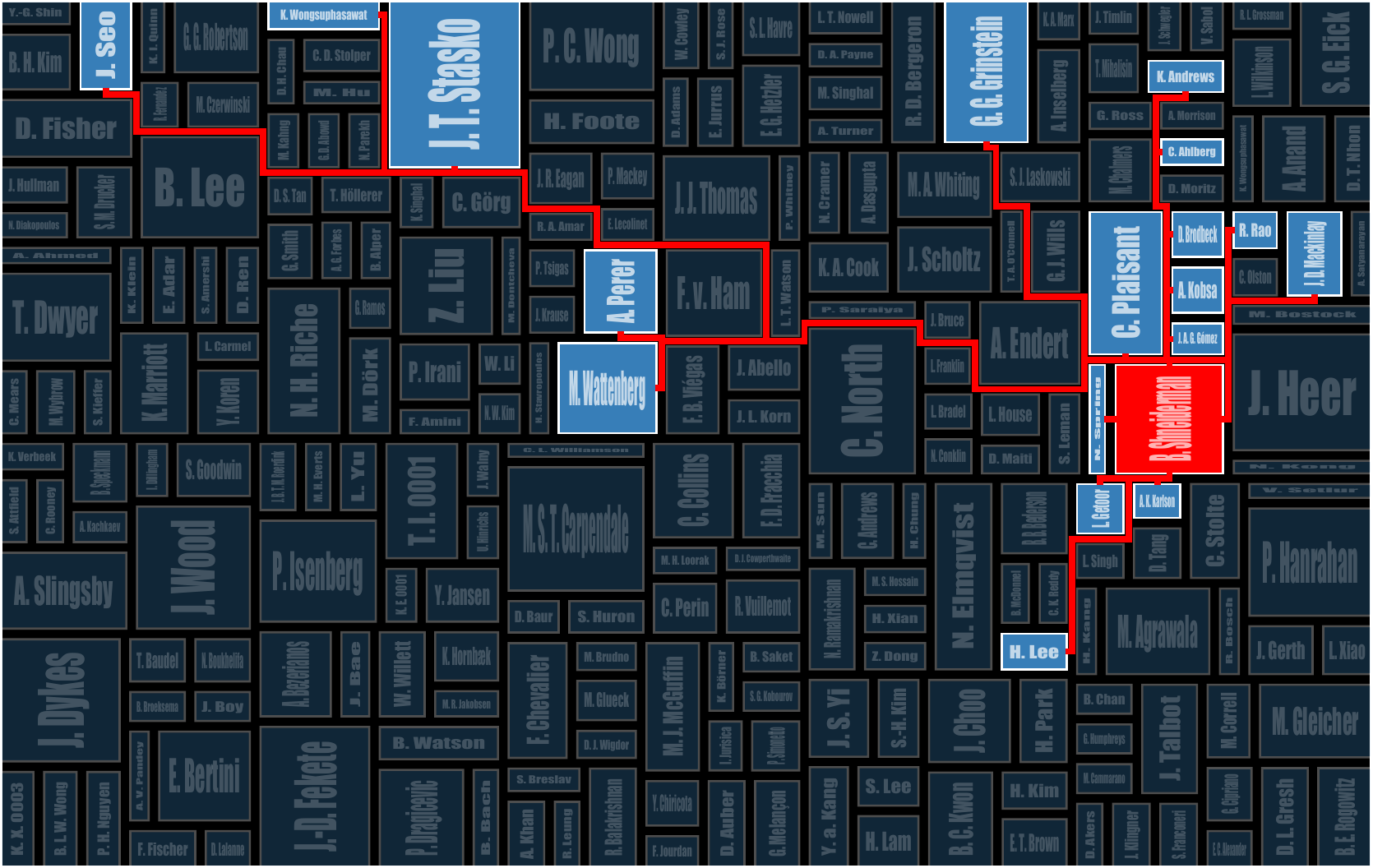}
   \caption{}
 \end{subfigure}
 
  \caption{(a): The green cluster in \autoref{fig:teaser}, indicating senior researchers are surrounded by many related young authors. (b): We hover \textit{Ben Shneiderman}, who is a pioneer of rectangular space-filling visualizations. His related authors are highlighted and connected by red channels. }
  \label{fig:vis_pub_case}
\end{figure*}

\section{Evaluation}

We evaluate the desired-aspect-ratio strategy on synthetic data, demonstrate the optimization effectiveness on three real-world datasets, and provide three use cases of VMap.

\begin{table}[!t]
\renewcommand{\arraystretch}{1.3}
\caption{Evaluation of aspect ratio quality for two algorithms: \textbf{s}caled \textbf{e}qual-\textbf{w}eight (SEW) partitioning strategy and our \textbf{d}esired-\textbf{a}spect-\textbf{r}atio (DAR) partitioning strategy on synthetic data. } \label{table:aspect_ratio}
\centering
\begin{tabular}{|c|c|c|c|}
\hline
Target Aspect Ratio & Points & Algorithm & $\textrm{loss}_{\textrm{r}}$ \\
\hline
\multirow{4}{*}{$\frac{3}{2}$}
      & \multirow{2}{*}{$10$}     & SEW    & $1.3234 \pm 1.1477$\\
      &                           & DAR     & $\bm{1.1605 \pm 1.2761}$\\
      \cline{2-4}  
      & \multirow{2}{*}{$100$}    & SEW    & $1.3303 \pm 0.3995$\\
      &                           & DAR     & $\bm{1.1111 \pm 0.3408}$\\
      \cline{2-4}    
      & \multirow{2}{*}{$1,024$}  & SEW    & $1.3151 \pm 0.1263$\\
      &                           & DAR     & $\bm{1.0989 \pm 0.1080}$\\
\hline 
\multirow{4}{*}{$\frac{1+\sqrt{5}}{2}$}
      & \multirow{2}{*}{$10$}     & SEW    & $1.3950 \pm 1.2158$\\
      &                           & DAR     & $\bm{1.1422 \pm 1.1126}$\\
      \cline{2-4}  
      & \multirow{2}{*}{$100$}    & SEW    & $1.4022 \pm 0.4322$\\
      &                           & DAR     & $\bm{1.1170 \pm 0.3449}$\\
      \cline{2-4}   
      & \multirow{2}{*}{$1,024$}  & SEW    & $1.3877 \pm 0.1292$\\
      &                           & DAR     & $\bm{1.1152 \pm 0.1076}$\\
\hline
\end{tabular}
\end{table}

\subsection{Study of Desired-Aspect-Ratio Partitioning Strategy}

Given a desired aspect ratio, We evaluate whether our DAR partitioning method can produce rectangles with good aspect ratios by comparing it with an improved variant of the equal-weight partitioning strategy. 

\revision{\textbf{Reasons of the baseline selection:} 
When the input is a set of points with 2D positions, the equal-weight strategy was used by \cite{heilmann2004recmap, duarte2014nmap, ghoniem2015weighted} to approach squares when subdividing rectangles. Duarte et al. \cite{duarte2014nmap} reported that the equal-weight strategy produces the state-of-the-art results for $1:1$ aspect ratio by a quantitative comparison with the spatially ordered treemap \cite{wood2008spatially}, the one-dimensional ordered treemap \cite{bederson2002ordered}, and the equal-number strategy \cite{mansmann2007visual, duarte2014nmap}. 
}

\revision{
\textbf{Scaled equal-weight (SEW) strategy:} 
For a fair comparison, we improve the original equal-weight strategy to generate rectangles approaching a desired aspect ratio $r$. The original equal-weight strategy of Nmap~\cite{duarte2014nmap} targets the generation of squarish rectangles, namely, the target aspect ratio of the equal-weight strategy in~\cite{duarte2014nmap} is $1:1$. To improve it, we first select the display space to be $1:1$. After we generate rectangles using the original equal-weight strategy. We scale the widths of both the display space and the produced rectangles by $r$, so that the resulting rectangles are expected to approach rectangles with the desired aspect ratio $r$. }

\revision{
Following previous treemap studies~\cite{bederson2002ordered, shneiderman2001ordered}, we study different methods by using Monte Carlo trials. We test two aspect ratios. Besides the $\frac{2}{3}$ suggested by Kong et al. \cite{kong2010perceptual}, we also test the golden ratio (i.e., $\frac{1+\sqrt{5}}{2}$), which is suggested by D3.js \cite{bostock2011d3, bostock2019d3url} for treemaps. We test three numbers of points ($10$, $100$, and $1,024$) to examine whether the point counts affect the results. Given an aspect ratio and a point count, we generate $10,000$ trials for measurement. For each trial, we synthesize points by generating locations of points randomly; the value of each dimension is sampled from $[0, 1)$ with a uniform distribution. As suggested by previous treemap experiments \cite{bederson2002ordered, shneiderman2001ordered}, weights of points are drawn from a log-normal distribution, whose underlying normal distribution is with a mean of $0$ and standard deviation of $1$. }


Results of experiments on \autoref{table:aspect_ratio} report the losses of aspect ratio with standard deviations across different experiment settings. 
The results demonstrate that our method surpasses the scaled equal-weight strategy by an average $17.22\%$ of relative loss reduction with respect to the aspect ratio loss. Also, when the number of points increases, the average losses of aspect ratio reduce a little, and the standard deviations decrease a lot for both methods. 

\begin{table*}[!t]
\renewcommand{\arraystretch}{1.3}
\caption{Evaluation of areal and topological errors for different methods. Statistics of CURDE and ECPA come from \cite{carrizosa2018mathematical}. ``VMap (Mean)'': we repeat the simulated annealing process $5$ times and record average statistics of the $5$ repetitions with standard deviations. ``VMap (Best)'': we report the best configuration and the total time of the $5$ repetitions. } \label{table:optimization}
\centering
\begin{tabular}{|c|c|c|c c c|c|c|}
\hline
Data & Algorithm & $\textrm{error}_{\textrm{a}}$ & Lost Edge & Fake Edge & $\textrm{error}_{\textrm{t}}$ & Total Error & Time (in minute) \\
\hline

\multirow{4}{*}{Netherlands}
      & CRUDE                       & $41.60\%$     & $6$      & $5$       & $40.74\%$     & $82.34\%$ & $180$ \\
      \cline{2-6} \cline{7-8} 
      & ECPA                        & $12.20\%$     & $0$       & $3$      & $12.00\%$     & $24.20\%$ & $21$ \\
      \cline{2-6} \cline{7-8} 
      & VMap (Mean)             & $2.41\%\pm{4.13\%}$      & $1.4\pm{0.8944}$       & $1.6\pm{0.8944}$      & $12.52\%\pm{6.63\%}$     & $14.93\%\pm{6.53\%}$      & $\bm{1.0703 \pm0.0519}$ \\
      \cline{2-6} \cline{7-8} 
      & VMap (Best)               & $0.00\%$      & $1$       & $1$      & $8.70\%$      & $\bm{8.70\%}$ & $5.3514$ \\      
\hline 
\hline
\multirow{4}{*}{Germany}
      & CRUDE                       & $43.80\%$     & $14$      & $14$      & $66.67\%$       & $110.47\%$ & $180$ \\
      \cline{2-6} \cline{7-8} 
      & ECPA                        & $29.00\%$     & $0$       & $7$       & $20.00\%$       & $49.00\%$  & $27$ \\
      \cline{2-6} \cline{7-8} 
      & VMap (Mean)            & $0.58\%\pm{0.69\%}$      & $4.6\pm{1.1402}$       & $9\pm{0.7071}$       & $36.77\%\pm{2.53\%}$       & $37.35\%\pm{2.93\%}$  & $\bm{2.1258 \pm0.0374}$ \\
      \cline{2-6} \cline{7-8} 
      & VMap (Best)               & $0.01\%$      & $3$       & $9$       & $32.43\%$       & $\bm{32.44\%}$  & $10.6289$ \\
\hline
\hline
\multirow{4}{*}{Blood}
      & CRUDE                       & $46.80\%$     & $2$  & $0$  & $10.53\%$ & $57.33\%$      & $180$ \\
      \cline{2-6} \cline{7-8} 
      & ECPA                        & $7.20\%$      & $2$  & $0$  & $10.53\%$ & $17.73\%$      & $120$ \\
      \cline{2-6} \cline{7-8} 
      & VMap (Mean)             & $0.00\% \pm{0.00\%}$      & $3.4\pm{0.5477}$  & $0\pm{0}$  & $17.89\%\pm{2.88\%}$ & $17.89\%\pm{2.88\%}$      & $\bm{0.4820\pm0.0012}$ \\
      \cline{2-6} \cline{7-8} 
      & VMap (Best)               & $0\%$      & $3$  & $0$  & $15.79\%$ & $\bm{15.79\%}$ & $2.4102$ \\
\hline 

\end{tabular}
\end{table*}

\begin{table}[!t]
\renewcommand{\arraystretch}{1.3}
\caption{
Exemplification for trading off topological error against areal error and aspect ratio loss by controlling the topological weight $\lambda_{\textrm{t}}$. } \label{table:optimization_topology}
\centering
\begin{tabular}{|c|c|c c|c|c|c|}
\hline
\multirow{2}{*}{Data} & \multirow{2}{*}{$\lambda_{\textrm{t}}$} & Lost & Fake  & \multirow{2}{*}{$\textrm{error}_{\textrm{t}}$} & \multirow{2}{*}{$\textrm{error}_{\textrm{a}}$} & \multirow{2}{*}{$\textrm{loss}_{\textrm{r}}$} \\
 & & Edge & Edge & & &
\\
\hline

\multirow{2}{*}{Netherlands}
      & 0.99             & $0$    & $0$  
      & $0\%$ & $55\%$    & $0.47$ 
      \\
      \cline{2-7} 
      & 0.01             & $9$   & $10$ 
      & $59\%$& $4\%$      & $0.04$ 
      \\ 
\hline 
\hline
\multirow{2}{*}{Germany}
      & 0.99             & $0$    & $4$  
      & $13\%$ & $87\%$     & $0.05$ 
      \\ 
      \cline{2-7} 
      & 0.01             & $16$   & $22$ 
      & $76\%$ & $4\%$     & $0.06$ 
      
      \\ 
\hline 
\hline
\multirow{2}{*}{Blood}
    & 0.99             & $2$    & $0$  
    & $11\%$ & $34\%$ & $0.30$ 
    \\ 
    \cline{2-7} 
    & 0.01             & $6$    & $2$  
    & $38\%$ & $4\%$  & $0.05$ 
    \\

\hline
\end{tabular}
\end{table}

\subsection{Study of Simulated Annealing based Optimization} \label{sect:evaluation_optimization}
\revision{
We compare the performance of different methods on all three datasets that were tested in the state-of-the-art optimization paper~\cite{carrizosa2018mathematical} including Blood, Netherlands, and Germany data. For Blood data, vertices represent blood groups, and the weights of vertices are proportions of people of the blood groups in the United States of America. Two blood groups are connected if one blood group can donate blood to the other. 
Netherlands data is introduced and discussed in~\autoref{sect:app_netherlands}. For the Germany data, vertices are states of Germany, and weights are areas of the states; edges represent the adjacency of the states. The detailed data of the three vertex-weighted graphs can be found in~\cite{carrizosa2018mathematical}. }

\revision{
We programmed our method by JavaScript, and measured its performance on a computer with a 2.30 GHz Intel Core i9-9880H processor and 16 GB memory. 
The number of annealing stages $ns$ is set to $16,384$ across all experiments; the errors and losses generally can be further reduced with more stages but with more time. }

\revision{
\textbf{Study for areal and topological errors:} 
We evaluate whether our simulated annealing based optimization method can effectively and efficiently produce rectangular subdivisions for graphs with low areal and topological errors by comparing with two state-of-the-art baselines in~\cite{carrizosa2018mathematical} including CRUDE and ECPA. Carrizosa et al.~\cite{carrizosa2018mathematical} programmed CRUDE and ECPA by AMPL (a mathematical programming language)~\cite{fourer1990modeling} and tested their performance on a computer with a 3.40 GHz Intel Core i7-2600K processor and a 16 GB memory. Because the methods in \cite{carrizosa2018mathematical} only focused on minimizing areal and topological errors, to make a fair comparison, we set the three weights to be $\lambda_{\textrm{a}}=0.5$, $\lambda_{\textrm{t}}=0.5$, and $\lambda_{\textrm{r}}=0$ for the loss function in our optimization. 
Also, we use \autoref{equa:error_topology} for topological error computation. 
Results of the experiments are shown on \autoref{table:optimization}. Our algorithm outperforms the two baselines on both total errors and optimization time across all three real-world datasets. 
}

\textbf{Study for trading-off topological error against areal error and aspect ratio loss: }
We set topological weight $\lambda_{\textrm{t}} \in \{0.01, 0.99\}$ for the optimization. The weights of areal error and aspect ratio loss are the same, namely, $\lambda_{\textrm{a}}=\lambda_{\textrm{r}}=\frac{1-\lambda_{\textrm{t}}}{2}$. 
We exemplify the results found by our simulated annealing based optimization on~\autoref{table:optimization_topology}. The results demonstrate that our method can make a trade-off among the three aspects by controlling their corresponding weights. 
Our method can improve topological performance by increasing the weight of topological error, $\lambda_{\textrm{t}}$, while, with the sacrifice of areal encoding and aspect ratio quality. This brings flexibility when users have different requirements for topological encoding, areal encoding, and aspect ratio on distinct applications, and our method is flexible to offer layouts with different trade-offs by specifying the weights for different errors and losses.

\subsection{Case Studies}

\revision{To perform vertex-centric graph exploration using VMap, we first identify vertices of interest based on their weights; larger weights are considered to have more potential interests in this design. Then, we can explore direct connections for vertices of interest and query other relationships by interactions. We demonstrate the usefulness of VMap in three applications, including visualization of the social network of characters in a novel, representation of academic communities, and generation of cartograms. }

\subsubsection{Social Network: \textit{Les Mis\'{e}rables} Data} \label{sect:app_les}
\textit{Les Mis\'{e}rables} dataset describes the co-occurrence network of characters in the novel \textit{Les Mis\'{e}rables} \cite{hugo1887miserables}, of which the data is provided in \cite{knuth1993stanford}. In this network data, each vertex is a character and has a weight that is the total of co-occurred times with other characters in the same section; higher co-occurrence counts of a character usually mean that the character is active in the novel. Two characters have an edge if they co-appear in the same section of the novel.

\revision{
In \autoref{fig:les-miserables-border-placeholder}d, VMap highlights important characters well such as \textit{Valjean}, \textit{Myriel}, and \textit{Marius}, which are centers of the purple, pink, and green clusters; also VMap preserves their main relationships well. \textit{Valjean} is the protagonist of this novel, who has been a prisoner for 19 years since stealing breads for his starving family members and does not trust laws; on the opposite, the connected character, \textit{Javert}, is a capable police inspector and pursues \textit{Valjean} as \textit{Valjean} escapes from the prison. \textit{Myriel} is a kind Bishop and a representation of mercy, and he saves \textit{Javert} and leads him to be an honest man; the related character \textit{Baptistine} is the sister of \textit{Myriel} and serves a realization of ``respectable''. \textit{Marius} is the second protagonist of this novel, and is the center of his family including his lover \textit{Cosette}, grandfather \textit{Gillenormand}, and aunt \textit{Mlle. Gillenormand}, in addition his neighbor, \textit{Thenardier} in \autoref{fig:les-miserables-border-placeholder}d. \textit{Marius} attends the society of \textit{Friends of the ABC} for the republican revolution of French. Most of the characters in the blue cluster are members of society and have mutual relationships with each other. }

\subsubsection{Academic Community: VIS Publication Data}
To test the scalability of our method, we use the VMap to visualize the co-authorship in the VIS publication data \cite{isenberg2017vispubdata}. The data records papers that were published in the VIS conference before $2016$. We display the authors who published at least two papers, for which $1,488$ authors are shown in \autoref{fig:teaser} by using VMap. As shown by \autoref{fig:vis_pub_case}a, we usually can observe that some professors are surrounded by his/her graduate students. Moreover, highly correlated research groups are placed closely. In \autoref{fig:vis_pub_case}b, we hover an author \textit{Ben Shneiderman} to observe his related researchers, where the relevant researchers are placed closely to \textit{Shneiderman}. 



\begin{figure}[htb]
\centering

 \begin{subfigure}[b]{0.45\columnwidth}
   \includegraphics[width=\columnwidth]{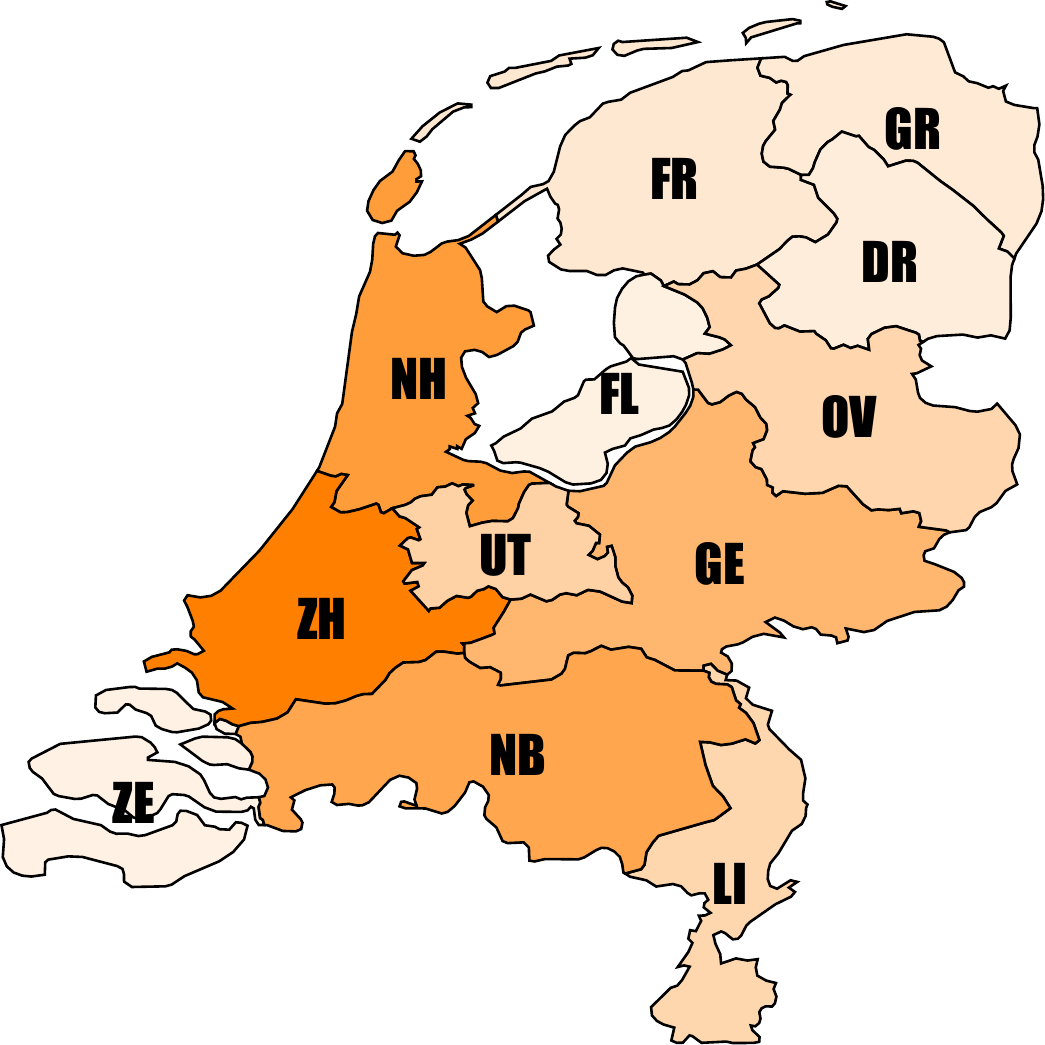}
   \caption{}
 \end{subfigure}
 \hfill
 \begin{subfigure}[b]{0.45\columnwidth}
   \includegraphics[width=\columnwidth]{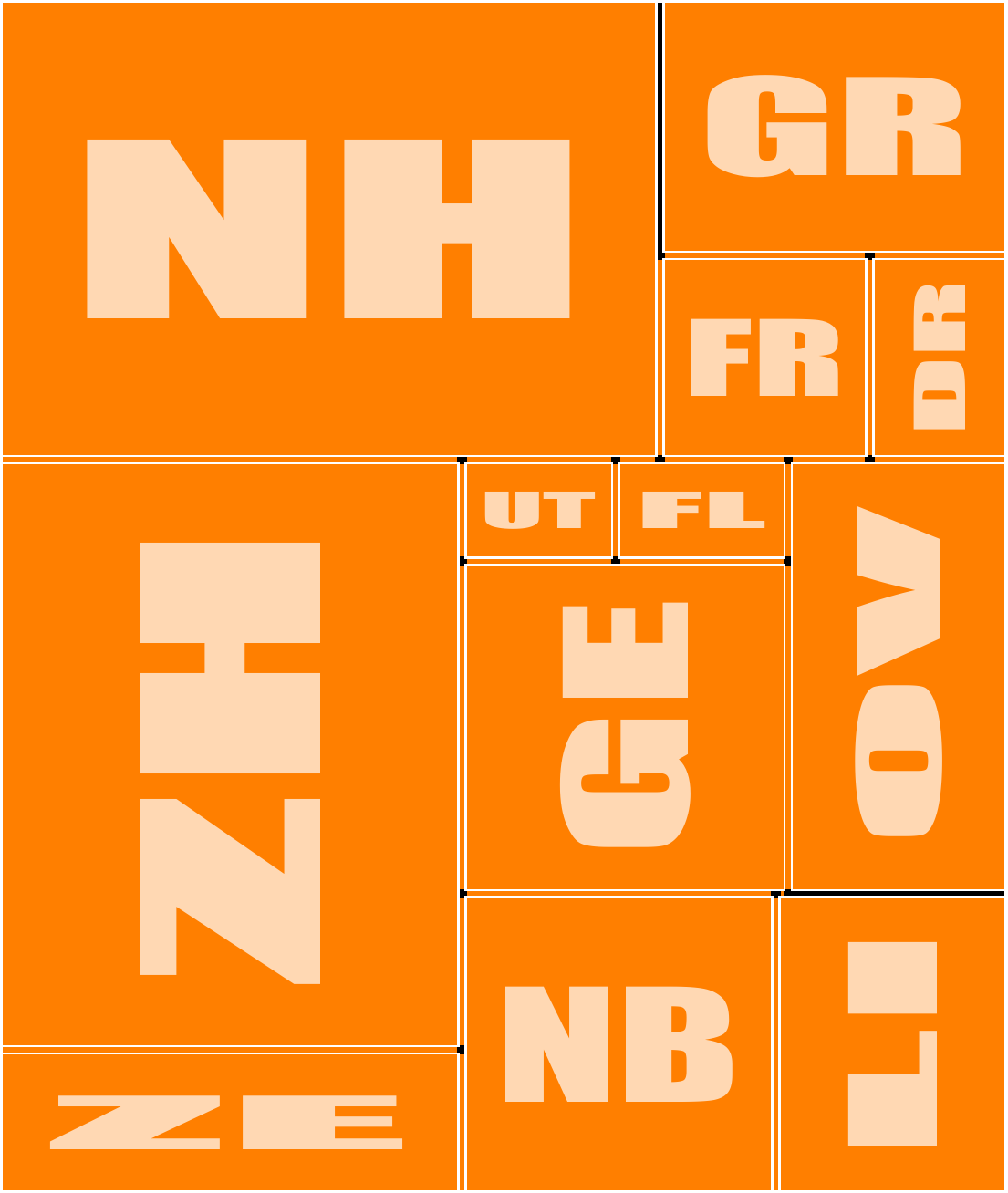}
   \caption{}\label{fig:netherlands_b}
 \end{subfigure} 

  \caption{\revision{Cartograms for provinces in the Netherlands. Labels on the cartograms are name abbreviations of corresponding provinces. Weights are the population of the provinces. 
  (a) A traditional choropleth map by displaying the population of different provinces using darkness. 
  (b) VMap encoding the population using areas of rectangles and preserves adjacency between provinces by using rectangular bridges; noted that the topological error (\autoref{equa:error_topology_amended}) is zero for this case.} }
  \label{fig:netherlands}
\end{figure}

\subsubsection{Cartogram: Netherlands Data} \label{sect:app_netherlands}
\revision{
On the Netherlands data, vertices are provinces of the Netherlands, and the weights of vertices are the population of the provinces. The edge of two provinces exists when the two provinces are contiguous geographically. \autoref{fig:netherlands} compares (a) a geographic map that uses darkness to encode the population and (b) a VMap that uses areas of rectangles for weight encoding. 
The results in \autoref{fig:netherlands}b indicate VMap preserves the adjacency of provinces well and can generate cartograms for weight analysis. 
}

\section{Discussions}

We discuss limitations and future works of VMap in the following. 

\revision{
\textbf{Limitation on representation of dense topologies:} 
Without interactions, static VMap uses bridges between rectangles encode direct connections between adjacent rectangles, which is useful for sparse graphs (e.g., pink and purple clusters of \autoref{fig:les-miserables-border-placeholder}d) or planar graphs (e.g., \autoref{fig:netherlands}). When the number of edges increases, topological structures with dense edges (e.g., complete graphs) may not be represented well by static VMap, which is a main limitation of the static VMap due to the limited adjacencies of rectangles. To remedy this, with the help of interactions, VMap supports query for the egocentric network of a selected rectangle and shorted-hop path between two selected rectangles. }

\textbf{Generation of treemap by using VMap:} 
Treemaps \cite{Shneiderman1992, johnson1991tree} are well-known for their ability to display hierarchical trees. Because a tree is a special form of a graph, we can also use our method to produce treemaps. To produce treemaps by our method, we treat each hierarchical tree as one graph and embed this graph into a 2D space.
After we acquire 2D embedding of the hierarchical tree, we use our partitioning based method to generate rectangles for vertices on the hierarchical tree layer by layer, which ensures rectangles of high-level vertices enclose low-level vertices. We will evaluate the effectiveness of treemaps generated by VMap in the future.

\section{Conclusion}

\revision{
This paper presents a novel rectangular space-filling design called VMap to visualize vertex-weighted graphs and perform vertex-centric graph exploration. VMap is effective in displaying information attached on vertices, including weights and labels. Also, VMap displays clusters of vertices in a compact way and places related individuals closely. The DAR partitioning method of VMap achieves a lower aspect ratio loss of resulting rectangles than the equal-weight partitioning based method for synthetic 2D spatial data. Moreover, when introducing the simulated annealing method to optimize the weighted loss of areal error, topological error, and aspect ratio loss, VMap outperforms existing optimization-based methods on total error and processing time on three real-world datasets. The border space is inserted between rectangles to route edges, which prevents vertex-edge occlusion. Static VMap visualizes edges by rectangular bridges between rectangles and is optimized to preserve as many direct contacts as possible. By interactions, VMap intuitively shows the egocentric network of a selected entity and displays the shortest-hop path between a pair of selected individuals. Finally, VMap is applied to three visualization applications, including social networks, academic communities, and cartograms, to demonstrate the efficacy. 
}


\bibliographystyle{abbrv}

\bibliography{vmap}
\end{document}



\firstsection{Two-Stage Rectangle Adjustment}

\maketitle



\begin{figure}[tb]
  \centering 
  \includegraphics[width=1.0\columnwidth]{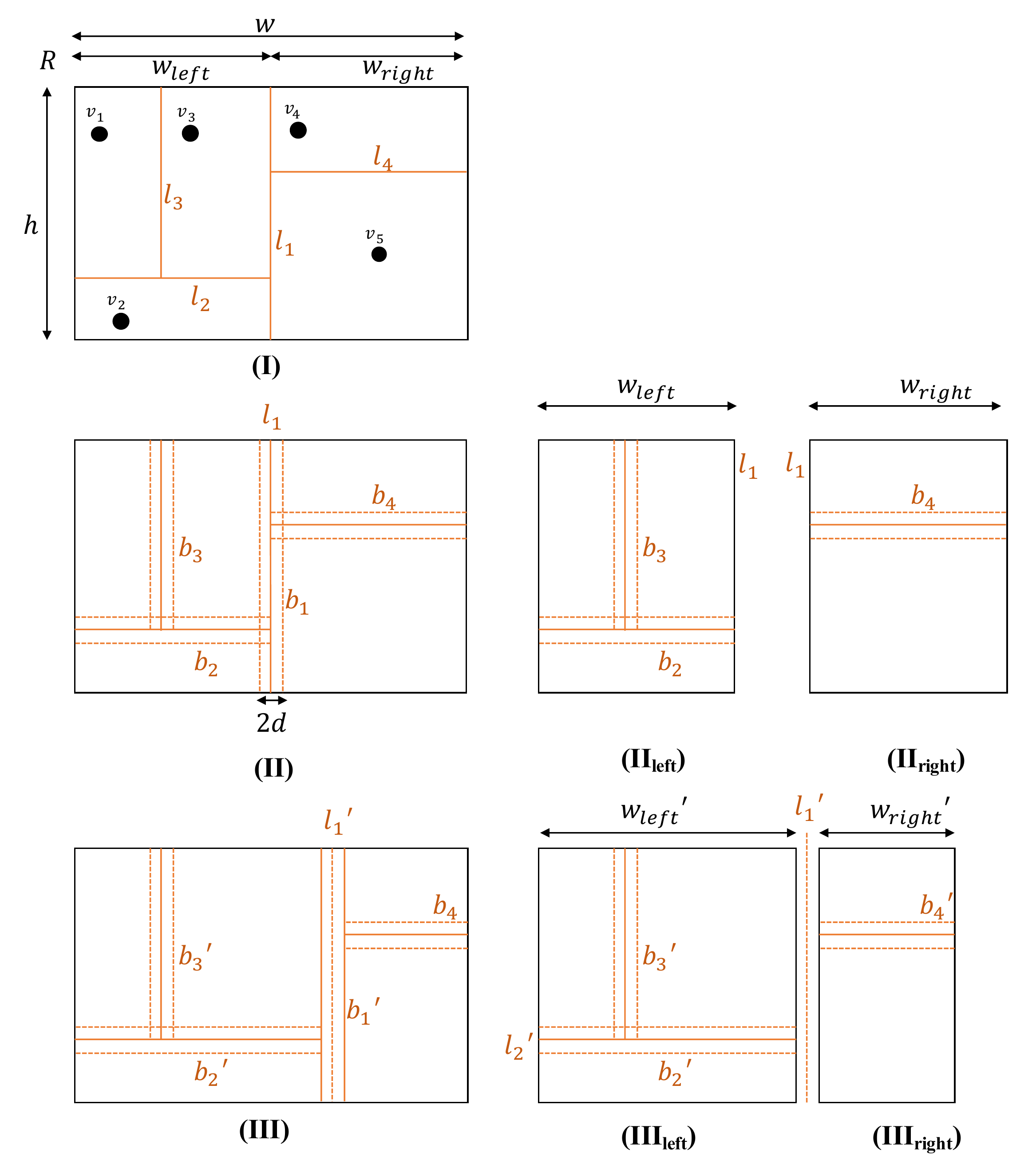}
  \caption{We illustrate how to adjust the subdivision of a rectangle $R$ with width $w$ and height $h$. (I) We have a rectangular partitioning of the $R$ by the order from $1$ to $4$, and the splitting lines are noted from $l_1$ to $l_4$. The five vertices from $v_1$ to $v_5$ have five corresponding leaf subdividing rectangles from $R_1$ to $R_5$. $l_1$ split $R$ into the left rectangle $R_{\textrm{left}}$ and the right rectangle $R_{\textrm{right}}$; the $R_{\textrm{left}}$ consists of $R_1$, $R_2$, and $R_3$; the $R_{\textrm{right}}$ consists of $R_4$ and $R_5$. (II) We expand the splitting lines to bands from $b_1$ to $b_4$ with widths of $2d$ to add borders for leaf rectangles. We display left and right parts of (II) on (II$_{\textrm{left}}$) and (II$_{\textrm{right}}$). However, the expansion of splitting lines distorts the encoding areas to encoding weights. (III) We adjust $l_1$ to $l_1'$ to correct the areal encoding. The gaps between (III$_{\textrm{left}}$) and (III$_{\textrm{right}}$) are placeholders for borders of leaf rectangles. }
  \label{fig:top-down-adjustment}
\end{figure}

\begin{figure}[tb]
  \centering 
  \includegraphics[width=1.0\columnwidth]{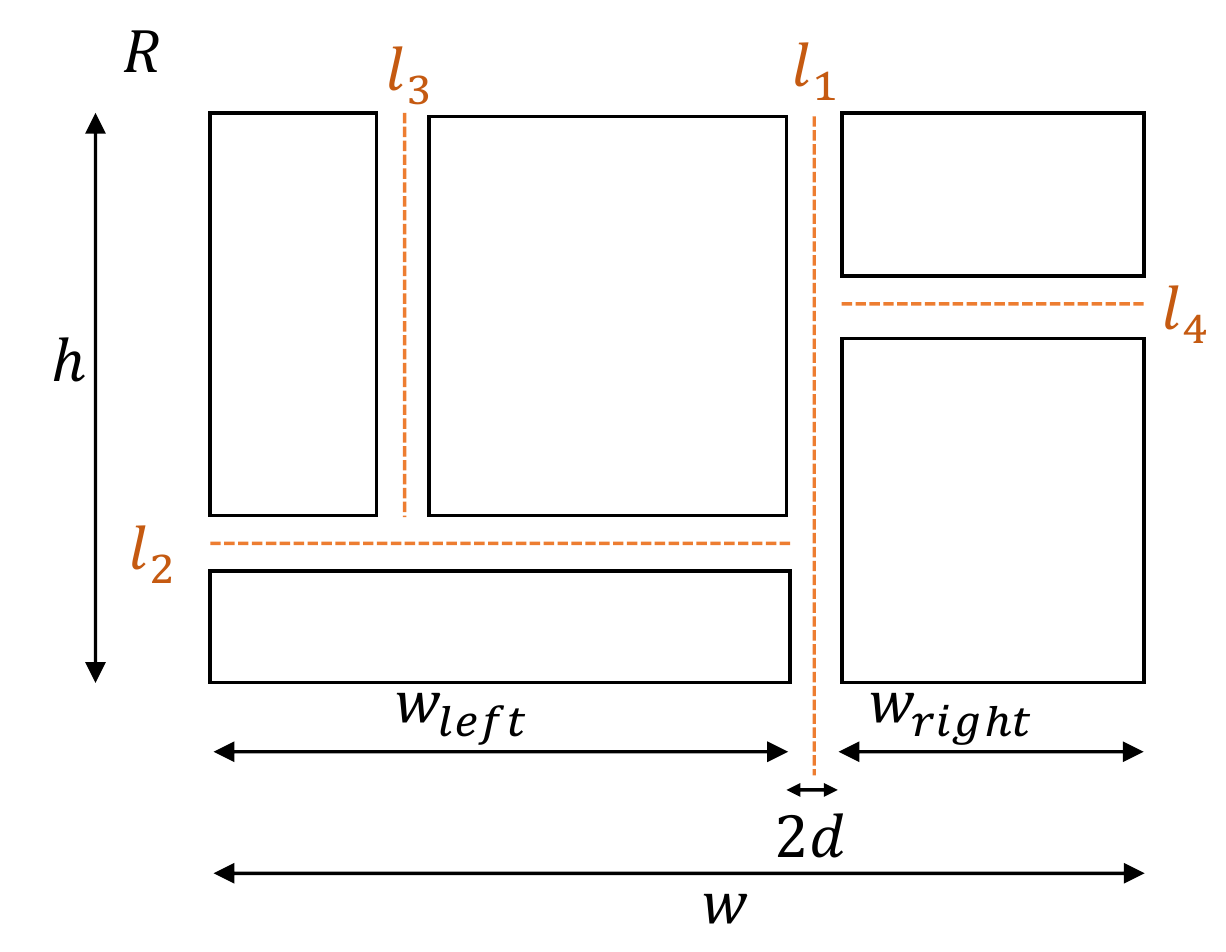}
  \caption{After the top-down adjustment, we have placeholders for borders of leaf rectangles. Next, we adjust the subdivisions bottom-up to ensure the faithful areal encoding. }
  \label{fig:bottom-up-adjustment}
\end{figure}

We give detailed computation derivations for both top-down and bottom-up adjustments and provide formal proof for proportion preservation as follows. 

\textbf{Coarse top-down adjustment:} 
We traverse the k-d tree of the given partitioning from the root to the leaves to scale the subdividing rectangles for two purposes: (1) adding a border with a fixed width $d$ for each leaf subdividing rectangle, and (2) balancing the areas of two subdividing rectangles of internal k-d tree nodes based on their given areal proportions.


Given a k-d tree node whose rectangle is $R$ with width $w$ and height $h$, we illustrate how to adjust its corresponding subdivision by the top-down adjustment method for horizontal partitioning in \autoref{fig:top-down-adjustment}; vertical partitioning is similar. 
We denote the left and right rectangles splitted by $l_1$ in \autoref{fig:top-down-adjustment} (I) by $R_{\textrm{left}}$ and $R_{\textrm{right}}$, respectively, and we adjust the two rectangles hereafter. 

Intuitively, the insertion of borders with a fixed width of $d$ is equivalent to expanding the inside splitting lines to bands with widths of $2d$ recursively, compared \autoref{fig:top-down-adjustment} (I) and (II)). 
As illustrated on \autoref{fig:top-down-adjustment} (II), we first expand splitting line $l_1$ into a band $b_1$. However, when the $b_1$ is inserted, the encoding areas of rectangles change. 
The \textit{encoding area} of a rectangle $R$ for areal encoding is equal to the area of the rectangle $R$ reducing the border area inside $R$, namely, 
\begin{equation} \label{equa:valid_area}
  \textrm{EncodingArea}_R = \textrm{Area}_R - \textrm{BorderArea}_R, 
\end{equation}
where the $\textrm{BorderArea}_R$ indicates the total border areas inside $R$. To compute $\textrm{BorderArea}_R$, we just calculate the area of the union of all bands on $R$, which is equal to the total lengths of splitting lines times $2d$ deducting by the intersection area between splitting bands:  
\begin{equation} \label{equa:boder_area}
\begin{aligned}
  & \textrm{BorderArea}_R \\
  = & 2d \cdot \textrm{SplittingLines} - \textrm{IntersectionArea} \\
  = & 2d \cdot \sum_{i=1}^m l_i - 2d^2 \cdot \textrm{JunctionNum}, 
\end{aligned}
\end{equation}
where $m$ is the number of splitting lines on $R$, and the $\textrm{JunctionNum}$ represents the number of junctions between splitting bands on $R$ and can be acquired by searching the k-d tree downward. 

To ensure that the encoding areas of $R_{\textrm{left}}$ and $R_{\textrm{right}}$ are proportional to their original areal proportions, we adjust the position of $l_1$ to $l_1'$ in \autoref{fig:top-down-adjustment} (III). Note that the number of junctions of subdividing rectangles remain unchanged if we adjust the position of $l_1$, for example, by comparing  \autoref{fig:top-down-adjustment} (II$_{\textrm{left}}$) and (III$_{\textrm{left}}$). But the total lengths of splitting lines change; for example, the length of $l_2$ on \autoref{fig:top-down-adjustment} (II$_{\textrm{left}}$) is enlarged on (III$_{\textrm{left}}$), and $l_4$ becomes shorter to $l_4'$. Hence, to decide the new position $l_1'$, we need to consider the change of the splitting lines. 

On \autoref{fig:top-down-adjustment}, we assume $w_{\textrm{left}}' = c_{\textrm{left}} \cdot w_{\textrm{left}}$ and $w_{\textrm{right}}' = c_{\textrm{right}} \cdot w_{\textrm{right}}$. To compute the values of $c_{\textrm{left}}$ and $c_{\textrm{right}}$, we have two equations. First, the sum of $w_{\textrm{left}}'$ and $w_{\textrm{right}}'$ should be equal to $w-2d$, hence
\begin{equation} \label{equa:linear_equation_1}
  w-2d = w_{\textrm{left}}' + w_{\textrm{right}}' = c_{\textrm{left}} \cdot w_{\textrm{left}} + c_{\textrm{right}} \cdot w_{\textrm{right}}
\end{equation} 
Second, the proportion between the new encoding areas $\textrm{EncodingArea}_{\textrm{left}}'$ and $\textrm{EncodingArea}_{\textrm{right}}'$ should be the same as the proportion between their original given areal proportions
\begin{equation} \label{equa:linear_equation_2}
  \frac{\textrm{EncodingArea}_{\textrm{left}}'}{\textrm{EncodingArea}_{\textrm{right}}'} = \frac{\sum_{i \in left}\alpha^P_i}{\sum_{j \in right}\alpha^P_j}
\end{equation}
where $\textrm{EncodingArea}_{\textrm{left}}'$ and $\textrm{EncodingArea}_{\textrm{right}}'$ can be represented by $c_{\textrm{left}}$ and $c_{\textrm{right}}$ respectively. Since we have two unknowns (i.e, $c_{\textrm{left}}$ and $c_{\textrm{right}}$) and two above equations, we can solve the system of two linear equations to acquire the values of $c_{\textrm{left}}$ and $c_{\textrm{right}}$. After that, the new position of $l_1$ is determined. The remaining problems are how to represent $\textrm{EncodingArea}_{\textrm{left}}'$ and $\textrm{EncodingArea}_{\textrm{right}}'$ by $c_{\textrm{left}}$ and $c_{\textrm{right}}$ respectively. 

We demonstrate how to express $\textrm{EncodingArea}_{\textrm{left}}'$ on \autoref{fig:top-down-adjustment} (III$_{\textrm{left}}$), and the computation of $\textrm{EncodingArea}_{\textrm{right}}'$ is similar. Note that by \autoref{equa:valid_area}, $\textrm{EncodingArea}_{\textrm{left}}' = \textrm{area}_{R_{\textrm{left}}'} - \textrm{BorderArea}_{R_{\textrm{left}}'}$. $\textrm{area}_{R_{\textrm{left}}'}$ can be expressed by $\textrm{area}_{R_{\textrm{left}}'} = w_{\textrm{left}}' \cdot h = c_{\textrm{left}} \cdot w_{\textrm{left}} \cdot h$. To compute $\textrm{BorderArea}_{R_{\textrm{left}}'}$, because $R_{\textrm{left}}$ on (II$_{\textrm{left}}$) is expanded horizontally to $R_{\textrm{left}}'$ on (III$_{\textrm{left}}$), the horizontal splitting line(s) including $l_2$ are enlarged by multiplying $c_{\textrm{left}}$ to become $l_2'$, but the vertical splitting line(s) such as $l_3$ remains the same. Hence, by \autoref{equa:boder_area}, 
\begin{equation} 
\begin{aligned}
&\textrm{BorderArea}_{R_{\textrm{left}}'} \\
= & 2d \cdot c_{\textrm{left}} \cdot \textrm{HorizontalSplittingLines} \\
+ & 2d \cdot \textrm{VerticalSplittingLines} \\
- & 2d^2 \cdot \textrm{JunctionNum}
\end{aligned}
\end{equation}
For example, on \autoref{fig:top-down-adjustment} (III$_{\textrm{left}}$), $\textrm{BorderArea}_{R_{\textrm{left}}'} 
= 2d \cdot (l_2' + l_3') - 2d^2
= 2d \cdot c_{\textrm{left}} \cdot l_2 + 2d \cdot l_3 - 2d^2$. Now, $\textrm{BorderArea}_{R_{\textrm{left}}'}$ is expressed by using $c_{\textrm{left}}$, and similarly $\textrm{BorderArea}_{R_{\textrm{right}}'}$ can be expressed by $c_{\textrm{right}}$. 

However, the top-down adjustment solely is not sufficient since the adjustment of low-level subdivisions affects high-level subdivisions that have been adjusted. For example, note that if we adjust $l_2'$ on \autoref{fig:top-down-adjustment} (III$_{\textrm{left}}$), the encoding area of $\textrm{EncodingArea}_{R_{\textrm{left}}'}$ also changes. If we move $l_2'$ upward, the area of $b_3'$ is shrunken, hence, $\textrm{BorderArea}_{R_{\textrm{left}}'}$ becomes fewer and $\textrm{EncodingArea}_{R_{\textrm{left}}'}$ becomes larger. This indicates that the adjustment of low-level subdivisions of the k-d tree hampers the adjusted high-level subdivisions. Hence, we introduce the fine bottom-up adjustment afterwards to solve the problem.

\textbf{Fine bottom-up adjustment:} 
The bottom-up adjustment traverses the k-d tree from leaves to the root to scale subdividing rectangles to preserve the given areal proportions. 

Given the rectangle $R$ (e.g., \autoref{fig:bottom-up-adjustment} for horizontal partitioning and similar for vertical partitioning), we move its splitting line $l_1$ to make the encoding areas of leaf subdividing rectangles have the same proportions as the given areal proportions $\{\alpha^P_i\}_{i=1}^{n}$. On \autoref{fig:bottom-up-adjustment}, we assume $l_2$, $l_3$, and $l_4$ have been adjusted correctly by this fine adjustment; since it is bottom-up, now we adjust $l_1$. After the movement of $l_1$, we also assume that $w_{\textrm{left}}' = c_{\textrm{left}} \cdot w_{\textrm{left}}$ and $w_{\textrm{right}}' = c_{\textrm{right}} \cdot w_{\textrm{right}}$. Hence, we can use the same system of equations, \autoref{equa:linear_equation_1} and \ref{equa:linear_equation_2}, to acquire $c_{\textrm{left}}$ and $c_{\textrm{right}}$. However, note that, to compute encoding areas of a rectangle for this fine adjustment, we can directly traverse the k-d tree downward to acquire the total areas of leaf rectangles because the leaf rectangles have excluded the areas of borders. Assume $\textrm{EncodingArea}_{\textrm{left}}$ and $\textrm{EncodingArea}_{\textrm{right}}$ store the total areas of leaf rectangles on the left and right of $l_1$ respectively. After the adjustment of $l_1$ to $l_1'$, the left and right valid areas become $\textrm{EncodingArea}_{\textrm{left}}' = c_{\textrm{left}} \cdot \textrm{EncodingArea}_{\textrm{left}}$ and $\textrm{EncodingArea}_{\textrm{right}}' = c_{\textrm{right}} \cdot \textrm{EncodingArea}_{\textrm{right}}$. Next, we solve the system of linear equations with two unknowns (i.e., $c_{\textrm{left}}$ and $c_{\textrm{right}}$) to determine the new position $l_1'$. 

\textbf{Proof of proportion preservation:} 
\revision{
We prove after the bottom-up adjustment, the given areal proportions $\{\alpha^P_i\}_{i=1}^{n}$ are correctly preserved by areas of adjusted rectangles. We note areas of adjusted rectangles $R_1$, $R_2$, ..., $R_n$ are correct if and only if their areas and the given proportions have the same multi-part ratio, namely, $(\textrm{area}_1:\textrm{area}_2:...:\textrm{area}_n) = (\alpha^P_1:\alpha^P_2:...:\alpha^P_n)$. }

\revision{
We first consider situations for internal nodes of the adjusted k-d tree. Given an internal node of the k-d tree, we assume that its left subtree contains leaf rectangles $R_1, R_2, ..., R_k$ and the leaf rectangles have been adjusted correctly, namely, $(\textrm{area}_1:\textrm{area}_2:...:\textrm{area}_k) = (\alpha^P_1:\alpha^P_2:...:\alpha^P_k)$. Similarly, we assume that its right subtree has leaf rectangles $R_{k+1}, R_{k+2}, ..., R_n$, and $(\textrm{area}_{k+1}:\textrm{area}_{k+2}:...:\textrm{area}_n) = (\alpha^P_{k+1}:\alpha^P_{k+2}:...:\alpha^P_n)$. Note that the scaling operation of the bottom-up adjustment ensures $(\sum_{i=1}^{k} \textrm{area}_i : \sum_{i=k+1}^{n} \textrm{area}_i) = (\sum_{i=1}^{k} \alpha^P_i : \sum_{i=k+1}^{n} \alpha^P_i)$. }

\revision{
\begin{lemma} \label{lemma1}
Given $(\textrm{area}_1:\textrm{area}_2:...:\textrm{area}_k) = (\alpha^P_1:\alpha^P_2:...:\alpha^P_k)$, $(\textrm{area}_{k+1}:\textrm{area}_{k+2}:...:\textrm{area}_n) = (\alpha^P_{k+1}:\alpha^P_{k+2}:...:\alpha^P_n)$, and $(\sum_{i=1}^{k} \textrm{area}_i : \sum_{i=k+1}^{n} \textrm{area}_i) = (\sum_{i=1}^{k} \alpha^P_i : \sum_{i=k+1}^{n} \alpha^P_i)$, then $(\textrm{area}_1:\textrm{area}_2:...:\textrm{area}_n) = (\alpha^P_1:\alpha^P_2:...:\alpha^P_n)$. 
\end{lemma}
}

\begin{proof}
Since $(\textrm{area}_1:\textrm{area}_2:...:\textrm{area}_k) = (\alpha^P_1:\alpha^P_2:...:\alpha^P_k)$, we have, 
\begin{equation} \label{equa:lemma1}
(\forall i \in [1, k]) \frac{\textrm{area}_i}{\sum_{j=1}^{k} \textrm{area}_j} = \frac{\alpha^P_i}{\sum_{j=1}^{k} \alpha^P_j}. 
\end{equation}
Similarly, we have
\begin{equation} \label{equa:lemma2}
(\forall i \in [k+1, n]) \frac{\textrm{area}_i}{\sum_{j=k+1}^{n} \textrm{area}_j} = \frac{\alpha^P_i}{\sum_{j=k+1}^{n} \alpha^P_j}. 
\end{equation}
Hence, considering \autoref{equa:lemma1} and \ref{equa:lemma2}, we have
\begin{equation} \label{equa:lemma3}
\begin{aligned}
& (\textrm{area}_1:...:\textrm{area}_k:\textrm{area}_{k+1}:...:\textrm{area}_n) \\
= & (\alpha^P_1 \cdot \frac{\sum_{i=1}^{k} \textrm{area}_i}{\sum_{i=1}^{k} \alpha^P_i}:...:\alpha^P_k \cdot \frac{\sum_{i=1}^{k} \textrm{area}_i}{\sum_{i=1}^{k} \alpha^P_i} \\
& :\alpha^P_{k+1} \cdot \frac{\sum_{i=k+1}^{n} \textrm{area}_i}{\sum_{i=k+1}^{n} \alpha^P_i}:...:\alpha^P_{n} \cdot \frac{\sum_{i=k+1}^{n} \textrm{area}_i}{\sum_{i=k+1}^{n} \alpha^P_i}). 
\end{aligned}
\end{equation}
Also, from $(\sum_{i=1}^{k} \textrm{area}_i : \sum_{i=k+1}^{n} \textrm{area}_i) = (\sum_{i=1}^{k} \alpha^P_i : \sum_{i=k+1}^{n} \alpha^P_i)$, we have 
\begin{equation} \label{equa:lemma4}
\frac{\sum_{i=1}^{k} \textrm{area}_i}{\sum_{i=1}^{k} \alpha^P_i} = \frac{\sum_{i=k+1}^{n} \textrm{area}_i}{\sum_{i=k+1}^{n} \alpha^P_i}. 
\end{equation}
We import \autoref{equa:lemma4} into \autoref{equa:lemma3} and obtain
$$(\textrm{area}_1:\textrm{area}_2:...:\textrm{area}_n) = (\alpha^P_1:\alpha^P_2:...:\alpha^P_n). $$
\end{proof}


\revision{
\begin{theorem} \label{theorem1}
The bottom-up adjustment can adjust all rectangles correctly by given proportions $\{\alpha^P_i\}_{i=1}^{n}$. 
\end{theorem}
}

\begin{proof}
We use mathematical induction to prove this theorem. 

\textit{Base case. }
We consider boundary conditions. Given any two leaf rectangles $R_1$ and $R_2$ with the same parent node on the k-d tree, by using the bottom-up adjustment, we scale them to ensure $(\textrm{area}_1:\textrm{area}_2) = (\alpha^P_1:\alpha^P_2)$. 

\textit{Inductive step. }
Given any internal node of the k-d tree, based on Lemma \ref{lemma1}, the reached leaf rectangles of this internal node are all adjusted correctly by using the bottom-up adjustment. 

\textit{Conclusion. }
After the bottom-up adjustment for the root of the k-d tree, all leaf rectangles are adjusted correctly. 
\end{proof}


\bibliographystyle{abbrv}

\bibliography{vmap}